\documentclass[twocolumn,twocolappendix]{openjournal}

\setcitestyle{authoryear,round}
\usepackage[T1]{fontenc}
\usepackage{subcaption}
\usepackage{aecompl}
\usepackage{multirow}
\usepackage{newtxtext,newtxmath}
\usepackage{mdframed,bm}
\usepackage[T1]{fontenc}
\usepackage{ae,aecompl}
\usepackage{graphicx}	
\usepackage{amsmath}	
\usepackage{booktabs}
\usepackage{color}
\usepackage{enumitem}

\usepackage{multirow}
\usepackage[pdftex,dvipsnames]{xcolor}

\usepackage{hyperref}

\hypersetup{
    urlcolor=black,
    citecolor=blue,
    linkcolor=blue,
  }%

\usepackage{physics}
\usepackage{cleveref}
\usepackage{comment}
\usepackage{array}
\usepackage{slashed}
\usepackage[dvipsnames]{xcolor}

\usepackage{xspace}
\usepackage{xcolor}
\usepackage{float}
\restylefloat{table}

\usepackage{physics}
\usepackage[normalem]{ulem}

\definecolor{bostonuniversityred}{rgb}{0.8, 0.0, 0.0}

\newcommand{\metacal}{\textsc{Metacalibration}}

\usepackage{eso-pic}

\shorttitle{Weak lensing around LSBGs in DES}
\shortauthors{Chicoine, et al.\ (DES)}

\begin{document}

\AddToShipoutPictureBG*{%
  \AtPageUpperLeft{%
    \hspace*{18cm}%
    \raisebox{-10\baselineskip}{%
      \makebox[0pt][l]{\textnormal{DES-2023-0778}}
 
}}}%

\AddToShipoutPictureBG*{%
  \AtPageUpperLeft{%
    \hspace*{20.25cm} 
    \raisebox{-11\baselineskip}{%
      \makebox[0pt][r]{\textnormal{FERMILAB-PUB-23-380-PPD}}
}}}%

\label{firstpage}
\title{Weak Gravitational Lensing around Low Surface Brightness Galaxies in the DES Year 3 Data}
\author{N.~Chicoine,$^{1}$
J.~Prat,$^{1,2}$
G.~Zacharegkas,$^{3}$
C.~Chang,$^{1,3}$
D.~Tanoglidis,$^{3}$
A.~Drlica-Wagner,$^{1,4,3}$
D.~Anbajagane,$^{3}$
S.~Adhikari,$^{5}$
A.~Amon,$^{6}$
R.~H.~Wechsler,$^{7,8,9}$
% ...
\begin{center} (DES Collaboration) \end{center}
\begin{center} \textit{Author affiliations are listed at the end of the paper} \end{center}}

\begin{abstract}

We present galaxy-galaxy lensing measurements using a sample of low surface brightness galaxies (LSBGs) drawn from the Dark Energy Survey Year 3 (Y3) data as lenses. LSBGs are diffuse galaxies with a surface brightness dimmer than the ambient night sky. These dark-matter-dominated objects are intriguing due to potentially unusual formation channels that lead to their diffuse stellar component. Given the faintness of LSBGs, using standard observational techniques to characterize their total masses proves challenging. Weak gravitational lensing, which is less sensitive to the stellar component of galaxies, could be a promising avenue to estimate the masses of LSBGs. Our LSBG sample consists of 23,790 galaxies separated into red and blue color types at $g-i\ge 0.60$ and $g-i< 0.60$, respectively. Combined with the DES Y3 shear catalog, we measure the tangential shear around these LSBGs and find signal-to-noise ratios of 6.67 for the red sample, 2.17 for the blue sample, and 5.30 for the full sample. We use the clustering redshifts method to obtain redshift distributions for the red and blue LSBG samples. Assuming all red LSBGs are satellites, we fit a simple model to the measurements and estimate the host halo mass of these LSBGs to be $\log(M_{\rm host}/M_{\odot}) = 12.98 ^{+0.10}_{-0.11}$. 
We place a 95\% upper bound on the subhalo mass at $\log(M_{\rm sub}/M_{\odot})<11.51$. By contrast, we assume the blue LSBGs are centrals, and place a 95\% upper bound on the halo mass at $\log(M_\mathrm{host}/M_\odot) < 11.84$. 
We find that the stellar-to-halo mass ratio of the LSBG samples is consistent with that of the general galaxy population. This work illustrates the viability of using weak gravitational lensing to constrain the halo masses of LSBGs.\\

\end{abstract}

\section{Introduction} \label{sec:intro}
Low-surface-brightness galaxies (LSBGs) are diffuse and generally dark-matter-dominated galaxies that are colloquially defined by a surface brightness fainter than the ambient night sky. These galaxies reside in environments ranging from the isolated open field to massive galaxy clusters \citep{Bothun1997,McConnachie2012, Martin2013, Danieli2017, Cohen2018, Leisman2017, Prole2021, Bhattacharyya2023}. They cover a diverse range of sizes, from compact dwarf galaxies to luminous galaxies five times the size of the Milky Way \citep{Das2013, Kado-Fong2021, Greene2022}. As outliers in the correlation between galaxy size and luminosity, LSBGs serve as a crucial benchmark for evaluating models that extrapolate galaxy characteristics from cosmological principles. Recent interest in LSBGs has grown due to the critical role they may play in understanding the physics of galaxy evolution \citep{Thuruthipilly2023}. Though current estimates indicate that LSBGs contribute little to the observable universe's stellar mass density and luminosity (<10\%) \citep{Bernstein1995, Driver1999, Hayward2005, Martin2019}, they may account for a sizable fraction of the total number density of galaxies (30\% - 60\%) \citep{McGaugh1996, Bothun1997, ONeil2000, Minchin2004, Martin2019}. The faintness of these aptly-named galaxies makes direct observation challenging, but their extreme characteristics present an opportunity to probe the nature of dark matter and test current theories of galaxy evolution.

Two of the largest catalogs of LSBGs to date \citep{Greco2018, Tanoglidis2021} have recently been  produced with searches using Hyper Suprime-Cam \citep[HSC,][]{Aihara2022} and Dark Energy Survey \citep[DES,][]{DES2016} data. Previous work focused on rich environments \citep{Koda2015, Mihos2015, Munoz2015, Martinez-Delgado2016, vanderBurgh2016, Yagi2016, Lee2018},
where it was easier to measure distances, and thus examine physical characteristics. Nonetheless, studying the properties of such galaxies for more general samples can prove interesting. For instance, \citet{Greene2022} estimated the redshift distribution of the HSC LSBG sample and inferred the physical properties, such as stellar mass and size, of the population. We take a similar approach here, but focus on the larger DES sample. Moreover, we proceed beyond estimating stellar masses from photometric magnitudes and aim to use gravitational lensing to provide mass estimates of the dark matter halos surrounding LSBGs. An improved grasp of LSBG halo masses, combined with stellar mass estimates, can provide us with a handle on the LSBG stellar-to-halo-mass relation \citep[SHMR,][]{Du2020,Moster2010, Behroozi2010}. At low masses, some uncertainty still persists in understanding the SHMR for the general galaxy population \citep{Danieli2023, Munshi2021, Brook2014}. Nonetheless, comparing the SHMR of LSBGs to the generalized SHMR allows us to test how closely LSBG characteristics align with the broader galaxy population. 

Gravitational lensing presents a promising avenue for constraining the halo masses of LSBGs. Gravitational lensing occurs when light from a background object, such as a star or a galaxy, passes by the gravitational potential well of a foreground mass \citep[e.g.,][]{Bartelmann2010}. The foreground (lens) object perturbs the light from the background (source) object, creating a distorted image. \textit{Weak} lensing refers to the case where we measure this distortion statistically instead of from a single galaxy \citep{Bartelmann2017}. \textit{Galaxy-galaxy lensing} specifically refers to the weak lensing measurement for which we statistically measure the distortion of an ensemble of source galaxies around an ensemble of lens galaxies. This technique effectively allows us to map out the average profile of the dark matter halos that host the lens galaxies.

Weak gravitational lensing has tantalized researchers with its potential for studying satellite and dwarf galaxies \citep{Thornton2023}. \citet{Sifon2018} measured the lensing signal around a sample of low-redshift satellite galaxies and constrained their subhalo masses. Similarly, \citet{wang2024} utilized galaxy-galaxy lensing to measure the subhalo mass of satellite galaxies in the redMaPPer cluster catalog from SDSS Data Release 8. \citet{Sifon2021} used weak lensing to estimate the mass of ultra-diffuse galaxies (UDGs), a subset of LSBGs frequently found in massive galaxy clusters. On the other hand, halo mass estimates of LSBGs have been obtained using X-ray photometry, as in \citet{Kovacs2019}, or by relying on the scaling relation between the halo mass and the number of associated globular clusters \citep{Prole2019}. Finally, the Merian Survey plans to conduct high signal-to-noise measurements of galaxy-galaxy lensing around dwarf galaxies \citep{Luo2023}. Here, we attempt to measure and model the galaxy-galaxy lensing signal of the DES Year 3 LSBG sample \citep{Tanoglidis2021} using the shapes of background galaxies from the DES Y3 shape catalog \citep{Gatti2021}. This represents the first example of a constraint on LSBG masses using weak lensing. 

The paper is organized as follows. In Sec.~\ref{sec:data}, we overview the data products used in this work and estimate the redshift distribution of the LSBG samples. In Sec.~\ref{sec:measurements}, we present the weak lensing measurements. In Sec.~\ref{sec:modeling}, we describe our model and the fit to the measurements. We estimate the stellar mass of LSBGs and obtain results for the stellar-to-halo mass relation in Sec.~\ref{sec:SHMR}. In Sec.~\ref{sec:conclusion}, we present our conclusions.

Throughout the paper we assume cosmological parameters from \citet{Planck2016} with $\Lambda$CDM model parameters $\Omega_{m}=0.3809$, $\Omega_{\Lambda}=0.6910$, and $\Omega_{b} = 0.0486$, where $\Omega_{m}$ is the total matter density of the universe, $\Omega_{\Lambda}$ is the dark matter density of the universe, and $\Omega_{b}$ is the baryonic matter density of the universe, all at redshift $z=0$. 

\section{Data} \label{sec:data}

Our data is based on the Dark Energy Survey \citep[DES,][]{DES2016}. DES is an optical near-infrared survey that covers approximately 5000 $\mathrm{deg}^2$ of the southern Galactic sky in five different filters ($grizy$), collecting data from hundreds of millions of distant galaxies up to a redshift of 1.4. The survey utilizes the Dark Energy Camera \citep{Flaugher2015} on the 4-m Blanco Telescope at the Cerro Tololo Inter-American Observatory. Data was collected over 758 distinct nights of observation. Our measurements utilize lens and source galaxies from the first three years (Y3) of DES observations \citep{Sevilla-Noarbe2021}.

\subsection{Lens Sample: LSBGs} \label{sec:lens catalog}

We use a lens sample of LSBGs selected from the DES Y3 images and described in detail in \citet{Tanoglidis2021}. The sample was constructed with a series of cuts to reject artifacts and reduce imaging contamination. 
\citet{Tanoglidis2021} defined extended LSBGs as galaxies with a $g$-band effective radii of 
\begin{equation}
\mathrm R_{\mathrm{eff}}(g) >2.5~\arcsec,  
\end{equation}
and a mean surface brightness of 
\begin{equation}
\mu_{\mathrm{eff}}(g) >24.3 \frac{\mathrm{mag}}{\arcsec^2}.
\end{equation}
Following the analysis of \citet{Greco2018}, the sample was further color-restricted to the ranges: 
\begin{equation} \label{eq:colorcut1} -0.1 < (g-i) < 1.4, \end{equation} 
\begin{equation} \label{eq:colorcut2} (g-r)  \ge 0.7 \times (g-i)-0.4, \end{equation}
\begin{equation} \label{eq:colorcut3} (g-r) \le 0.7\times (g-i)+0.4, \end{equation}

\noindent to reduce false detections caused by optical artifacts and blends of high-redshift galaxies. 
Objects in the sample were required to have an ellipticity  $ \le 0.7 $ to remove high-ellipticity artifacts (i.e. diffraction spikes). This round of cuts resulted in a sample size of 419,985 objects. The sample was then passed through a machine learning classification algorithm to remove additional sources of contamination, such as faint objects blended in the diffuse light from bright stars, ejections of tidal material from high-surface-brightness host galaxies, or blurring from bright regions of Galactic cirrus \citep{Tanoglidis2021}. Applying the machine learning algorithm reduced the sample size to 44,979 objects. The objects were visually inspected as cutouts with the DESI Legacy Imaging Surveys sky viewer \citep{Dey2019} and fit with Sersic profiles, shrinking the sample size to 23,790 objects. The average angular number density of LSBGs in the \citet{Tanoglidis2021} sample is $4.5~\mathrm{gal}/\mathrm{deg}^2$.

Galaxy colors correlate with galactic stellar populations, morphologies, and environments. In a generic galaxy sample, galaxy color distributions are bimodally split into red and blue color categories \citep{Blanton2003}. In Fig.~5, \citet{Tanoglidis2021} shows that the LSBG sample photometry is also bimodal. Applying a selection at  $ g-i = 0.60$ divides the sample into red LSBGs ($g-i \ge 0.60$, 7,805 galaxies) and blue LSBGs ($ g-i < 0.60$, 15,985 galaxies). We follow the same division in this work. As seen in Fig.~8 of \citet{Tanoglidis2021}, the red galaxies are strongly clustered, while the blue galaxies are scattered across the field. The two populations have similar angular size distributions and median Sersic indexes. Their brightness distributions differ, with blue galaxies appearing noticeably brighter \citep{Tanoglidis2021}. 

The DES Y3 LSBG sample does not come with redshifts, which are needed to model the lensing signal. As such, we describe our procedure to estimate the ensemble redshift distributions below.

\subsubsection{Redshift Distribution Estimation for the Lens Sample}\label{sec:cross correlation measurements} 

Following the work of \citet*{Giannini2022}, we use the clustering redshifts method to obtain an estimate of the redshift distribution for our LSBG lens sample. Clustering-based redshift methods derive the redshift distribution of an ``unknown'' galaxy sample by exploiting the two-point correlation signal between the unknown sample and a ``reference'' sample of galaxies with trusted redshifts. This process assumes that the cross-correlation signal between the two samples is positive when the objects overlap in physical space. 

Here, we cross-correlate the positions of the LSBG catalog with the positions of the all-sky 2MASS Photometric Redshift Catalog \citep[2MPZ,][]{Bilicki2014}, which contains approximately $10^6$ galaxies. This catalog cross-matches the 2MASS Extended Source Catalog \citep{Jarrett2000}, WISE \citep{Kovacs2013}, and SuperCosmos \citep{Hambly2001} samples and utilizes the artificial neural network approach \citep[ANNz,][]{Collister2004} to derive the photometric redshifts \citep{Jarrett2000, Hambly2001, Kovacs2013} of the sampled galaxies. The catalog has a median redshift of $z \sim 0.1$, a maximum redshift of $z\sim0.4$, and a precision of $\sigma_z \sim 0.015$. \citet{Tanoglidis2021} compared the clustering properties of the 2MPZ sample to the LSBG catalog to understand the clustering of the DES galaxies as a function of surface brightness. They found that the 2MPZ sample exhibited less clustering at intermediate angular scales ($0.1^\circ\le \theta \le 4^\circ$).

The boost factor measurements depicted in Fig.~\ref{fig:boost_factor}, along with the findings from  \citet{Greco2018}, \citet{Tanoglidis2021} and \citet{Greene2022}, lead us to anticipate a low redshift range for the LSBG sample, likely $z\lesssim 0.2$. For cross-correlation with the red LSBG sample, we divide the reference sample into thin redshift bins with a width of 0.01 over the range [0.0-0.14], finding little cross-correlation signal beyond this range. We instead use bin widths of 0.02 over the range [0.0-0.20] for cross-correlation with the blue sample to produce a stronger cross-correlation signal and to account for the widened spread of the redshift distribution. We also use the higher-redshift \textsc{redMaGiC} sample \citep{Rozo2016, Pandey2022} and the Magnitude-Limited Sample \citep[\textsc{MagLim},][]{Porredon2022} to ensure that the LSBGs do not have an extended high redshift tail.  

The clustering redshift signal represents an integral over the product of each population's galaxy-matter bias with the dark matter density two-point correlation function (e.g., Eq.~9 from \citealt{Giannini2022}). Here, we use the procedure proposed by \citet{Menard2013} and \citet{Schmidt2013}, which assumes that the target population (the sample of LSBGs) spans a narrow redshift range such that the galaxy bias of both the target (LSBG) and tracer (2MPZ) samples can be ignored. \citet{Menard2013} further explained how clustering can still be used in the non-linear regime for this application. Under these assumptions, also adopted by \citet{Greene2022}\footnote{We note that \citet{Greene2022} adopted this methodology to estimate the redshift distribution of the HSC LSBG sample, but did not include the  $w_\mathrm{DM}$ correction factor.} for a similar sample of LSBGs,  the redshift distribution can be estimated at the central redshift of each bin $z_i$ as
\begin{equation}\label{eq: redshift distribution} 
n_u(z_i) \propto \frac{\overline{w}_{ur}(z_i)}{w_\mathrm{DM}(z_i)},
\end{equation}
where $u$ stands for the unknown sample, $r$ represents the reference sample, and $\overline{w}_{ur}(z_i)$ is the cross-correlation measurement averaged over a given set of angular scales via \begin{equation}\label{eq:weighted_cross_correlation} \overline{w}_{ur} = \int_{\theta_\mathrm{min}}^{\theta_\mathrm{max}} W(\theta)w_{ur}d\theta.
\end{equation} 
We use $W(\theta)\propto \theta^{-1}$ as a weighting function to yield optimal S/N in the presence of shot noise \citep{Giannini2022}.
$w_{\rm DM} (z_{i})$ represents the dark matter density two-point correlation function, which can be estimated analytically. We use the revised fitting function \textsc{Halofit} from \citet{Takahashi2012} to model the non-linear matter spectrum and obtain $w_\mathrm{DM}$.  We implement this model using the public version of \textsc{CosmoSIS} \citep{Zuntz2015}. 

Following Eq.~\ref{eq:weighted_cross_correlation}, we average these measurements over the following means of the angular scale bins: 
$\theta=[32.45, 45.23, 63.25]$ arcmin, which correspond to a physical scale range of $0.20 - 9.37$ Mpc.  We select these angular bins because they act as middle points between the smallest scales that adhere to the theoretical requirements of the $w_\mathrm{DM}$ term \citep{Takahashi2012} and larger scales with increased errorbars that may suffer from observational systematic effects. For comparison, \citet{Greene2022} used scales between 0.1 and 3 Mpc. 

We compute the cross-correlation signal as a function of angular scale by using the \citet{Davis1983} estimator: 
\begin{equation}\label{eq:davis-peebles} 
w_{ur}(\theta) = \frac{N_{Rr}}{N_{Dr}} \frac{D_u D_r (\theta)}{D_u R_r(\theta)} -1, 
\end{equation}
where $D_u D_r (\theta)$ and $D_u R_r (\theta)$ stand for the data-data and data-random pairs, and $N_{Dr}$ and $N_{Rr}$ correspond to the total number of galaxies in the reference sample and the reference random catalog, used for normalization. The sample of random points comes from the Y3 footprint catalog \citep{Sevilla-Noarbe2021}, contains 4,570,480 data points, and matches the footprint of the LSBG lens/unknown sample. We apply this same positional mask to the 2MPZ catalog and compute the cross correlation. 

\begin{figure}
    \centering
    \includegraphics[width=0.45\textwidth]{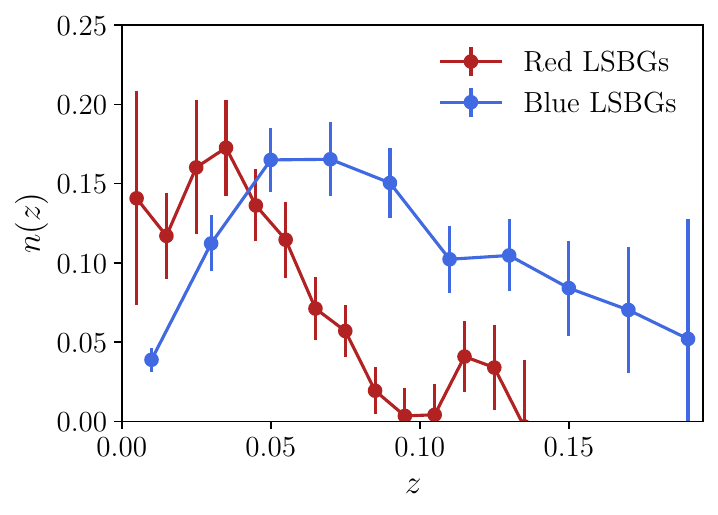}
    \caption{Normalized redshift distribution of the red and blue LSBGs averaged over the angular scales of $\theta=32.45, 45.23, 63.25$ arcmin. The errorbars are derived from the jackknife covariance of the total distribution. }
    \label{fig:redshift_distribution_avg}
\end{figure}

We perform this procedure for the red and blue LSBG samples and show the resulting redshift distributions in Fig.~\ref{fig:redshift_distribution_avg}. We obtain the errorbars by using the jackknife resampling method with 100 patches across the survey area. We assume no uncertainty comes from the theoretical input for $w_{\mathrm{DM}}$. We find that the red LSBG sample resides at lower redshifts than the blue LSBG sample. This difference may be influenced by the LSBG clustering behavior, as the majority of the red LSBG sample is associated with nearby structures (z<0.10) \citep{Tanoglidis2021}. In addition, blue LSBGs tend to have higher surface brightness, thus we may be able to detect them at higher redshifts.

\subsection{Source Sample}\label{sec:source catalog}

\begin{figure}
    \centering
    \includegraphics[width=0.45\textwidth]{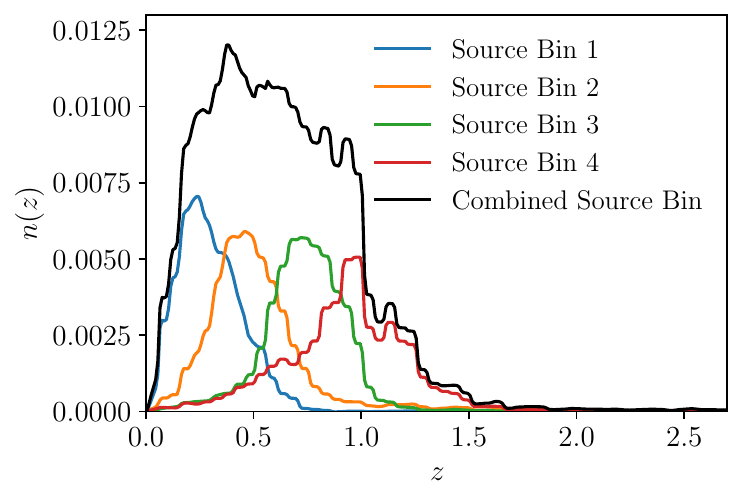}
    \caption{Redshift distribution of source galaxies, with a mean redshift of  0.6312 for the full sample. The four redshift bins are weighted by galaxy count, summed, and normalized to produce the combined source bin distribution.}
    \label{fig:source_galaxy_redshift}
\end{figure}

The background sources come from the DES Y3 \metacal\ shape catalog described in \citet*{Gatti2021}. This catalog utilizes the \metacal\ technique, developed by \citet{Sheldon2017}. Shear is measured from the ellipticity of the source galaxy image, but this initial measurement is noisy and biased. \metacal\ corrects this bias by applying a series of artificial shears in different directions to an image and evaluating the response of the ellipticity. The redshift distribution of this catalog and its uncertainties are calibrated with the framework described in \citet*{Myles2021}. This framework
utilizes a particular category of neural network, known as a self-organizing map (SOM), to study the relationship between observed galaxy colors and their redshifts. Once trained with large datasets of photometry for galaxies with known redshifts, the algorithm can estimate the redshifts of new galaxies based on their photometric data. The resulting redshift distribution shapes are further constrained using the same clustering method described in Sec.~\ref{sec:cross correlation measurements}. 
The Y3 \metacal\ shape catalog includes four tomographic redshift bins. To reduce the shape noise and boost the signal, we weight the redshift bins by their galaxy count and combine them, as shown in Fig.~\ref{fig:source_galaxy_redshift}. 
 Moreover, as the LSBGs live at low redshifts, the lensing efficiency varies little between the different redshift bins, suggesting that we gain no significant information by preserving their separation. In addition, combining the source bins at the measurement level circumvents the need to measure the cross-covariance between redshift bins.

\section{Tangential Shear Measurements} \label{sec:measurements}
We cross-correlate the shapes of background source galaxies with the positions of foreground lens galaxies.
The lensing distortion orients the source image tangentially around the lens image. For a particular lens-source galaxy pair (\textit{LS}), we find the tangential component of the ellipticity via:
\begin{equation} \label{eq:ellipciticy} 
e_{t,LS} = -e_1\cos(2\phi)-e_2\sin(2\phi).
\end{equation}
Here we define $\phi$ as the position angle of the source galaxy centered at the lens galaxy with respect to the declination-axis of the sky coordinate system, and $e_1$ and $e_2$ as the ellipticity components defined in the same sky coordinate system \citep{Prat2022}.

We average the tangential component of the ellipticity over many source-lens pairs 
to obtain our estimator $\gamma_{t}$. The final form of our estimator can be written as 
\begin{equation}\label{eq:finalshearestimator} 
\gamma_t (\theta) = \frac{1}{\left\langle{R}\right\rangle}\left[\frac{\Sigma_{LS}w_{LS}e_{t, LS}(\theta)}{\Sigma_{LS}w_{LS}(\theta)} - \frac{\Sigma_{RS}w_{RS}e_{t, RS}(\theta)}{\Sigma_{RS}w_{RS}(\theta)}\right],  
\end{equation}
with $w_{LS} = w_L w_S$ as the weight factor for a particular lens-source pair, where $w_L$ is the weight of the lens galaxy and $w_S$ is the weight of the source galaxy. $\theta$ is the angular separation between the lens and the source galaxy pair. In practice, we measure $\gamma_{t}$ in angular bins, thus here we sum over lens-source pairs whose separation falls into a given angular bin. The lens galaxy weights are uniform ($w_L$ = 1). The second $RS$ term represents the cross-correlation of the random-source pairs. This term is used to remove the spurious signal coming from the edges of the survey and masked regions \citep{Sevilla-Noarbe2021}. At large scales, applying random point subtraction reduces the covariance, as seen in \citet*{Singh2017, Prat2018}. $\left\langle{R}\right\rangle$ represents the response factor, the self-calibration term derived from the \metacal\ technique \citep{Sheldon2017}. We have validated our code by calculating the responses for each of the four tomographic source redshift bins and verifying that they match table 4 of \citet{Prat2022}. The response factor for the combined source redshift distribution is 0.7184.

\begin{figure}
    \centering
    \includegraphics[width=0.45\textwidth]{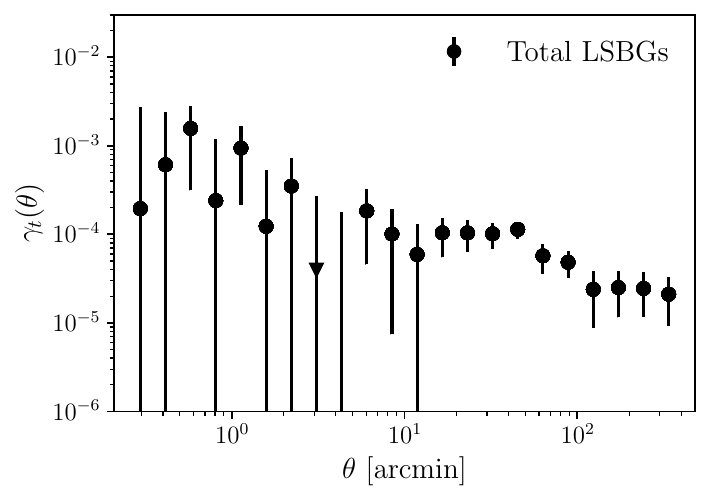}
    \caption{Tangential shear measurements for the full sample of LSBGs. The measurements span from 0.25-400 arcmin and are divided into 22 angular bins. We display the absolute value of the shear measurements -- the triangle marking indicates a bin with a negative shear measurement. The errorbars are derived from the jackknife covariance.}
    \label{fig:LSBG_measurements}
\end{figure}

The above estimator is implemented using the software package \textsc{TreeCorr} \citep{Jarvis2004}. 
We use the \texttt{NGCorrelation} class from \textsc{TreeCorr} to calculate the position-shape correlations. We establish the maximum scales of the measurements at 400 arcmin, as beyond this point the tangential shear signal becomes consistent with random noise. We carry out measurements in 22 angular bins spanned logarithmically from 0.25 to 400 arcmin, where the lower bound represents the cutoff for the DES Y3 galaxy-galaxy lensing measurements used in the small scale extension from \citet{Zacharegkas2022}. Based on the survey area $ (\sim5000~\mathrm{deg}^2$) and our largest measured angular scales, we split the lens galaxies and random points into 100 patches across the survey area to obtain the jackknife covariance. The tangential shear measurements for the total LSBG sample are displayed in Fig.~\ref{fig:LSBG_measurements}. The red and blue samples are shown in Fig.~\ref{fig:red_blue_galaxy_shear_model}.
We perform a series of diagnostic tests with this data vector in Appendix~\ref{sec:measurementvalidation} to ensure that the measurements are robust.

We calculate the $\chi^2$ statistics for the tangential shear measurements: 
\begin{equation} \label{eq:chi2}
\chi^2 = (\gamma_{t_d} - \gamma_{t_m})\cdot \textbf{C}^{-1} \cdot (\gamma _{t_d} - \gamma_{t_m})^\mathrm{T}, 
\end{equation}
\noindent where $\gamma _{t_d}$ is the measured tangential shear, $\gamma_{t_m}$ is the tangential shear model, and $\textbf{C}^{-1}$ is the inverse covariance. To correct for bias in the inverse covariance introduced by noise in the jackknife covariance, we multiply the inverse covariance by the Hartlap-Kaufman factor \citep{Kaufman1967, Hartlap2007}, 
\begin{equation}\label{eq:hartlap}
f = (n - m - 2) / (n - 1),
\end{equation} 
where $n$ is the number of realizations (100) and $m$ is the number of entries in the data vectors (22). 

To estimate the signal-to-noise for each data vector, we set $\gamma_{t_m}$ as a null signal and define the signal-to-noise as 
\begin{equation} \label{eq:SN} 
\mathrm{S/N} = \sqrt{\chi ^2_\mathrm{null} - \nu}, 
\end{equation} 
where $\nu$ corresponds to the number of degrees of freedom (22) \citep{Prat2022}. The signal-to-noise and $\chi ^2_\mathrm{null}$ values are listed in Table ~\ref{tab:signal-to-noise}. We find a weak signal for the blue LSBGs and a clear signal for the red LSBGs and the full LSBG sample. Despite containing fewer galaxies than the full sample, the red LSBG population produces the strongest shear signal, likely due to the clustered environments and the potential dilution of the combined sample's signal. We note that the blue and red samples produce similar lensing signals at smaller angular scales, $\theta\lesssim 20~\mathrm{arcmin}$.

\begin{table}
\centering
\begin{tabular}{ c c c c} 
 \hline
  & $\chi ^2_\mathrm{null}$/$\nu$ & S/N & Galaxy Count\\ 
 \hline
 Full Sample  & 50.09/22 &  5.30 & 23,790 \\ 
 \hline
 Red & 66.42/22 & 6.67 & 7,805
 \\
 \hline
 Blue & 26.70/22 &  2.17 & 15,985 \\
 \hline
\end{tabular}
\caption{Signal-to-noise and $\chi^2$ values for tangential shear measurements of red, blue, and total LSBG samples. $\nu$ represents the degrees of freedom. We note that for the modeling in Sec.~\ref{sec:modeling}, we only use the first fifteen tangential shear measurements around the blue LSBG sample.}
\label{tab:signal-to-noise}
\end{table}

Upon visually inspecting the measured lensing signal, we note a distinctive feature: the red sample appears to detect a mass distribution separated by some angular separation from the LSBGs, whereas the blue sample does not display this behavior. This observation aligns with prior expectations that the red LSBGs are primarily satellites of more massive galaxies \citep[e.g.,][]{Bhattacharyya2023}, but its confirmation via lensing is a novel insight, unveiling the dark matter environments surrounding these galaxies.

Indeed, Fig.~8 of \citet{Tanoglidis2021} shows that the red LSBG sample exhibits a preference for clustered environments. Furthermore, theoretical models suggest that faint, clustered, red galaxies are predominantly satellites of more massive dark matter halos. \citet{Berlind2005} utilized cosmological simulations to explore the relationship between galaxy color, luminosity, and environment, demonstrating that low-luminosity red galaxies primarily exist as satellites in massive halos. \citet{Zehavi2011} investigated the luminosity and color dependence of galaxy clustering in the SDSS Seventh Data Release (DR7), concluding that the strong clustering observed in faint red galaxies indicates their satellite status in massive halos.

Conversely, \citet{Thornton2023} found that samples of bluer galaxies at dwarf-galaxy scales reside preferentially in low-density regions and have low satellite fractions. Considering their scattered spatial distribution, we anticipate that the blue LSBG sample is likely comprised of central field galaxies. The lensing profile of these galaxies is consistent with the lack of a significant contribution from nearby massive objects.

Given these concurring results, we assume the red LSBGs are dominated by satellite galaxies and the blue LSBGs by central galaxies for the model fitting procedure described in the next section. Nevertheless, we test the validity of this assumption with alternative models in Appendix \ref{sec:hodmodels}.

\section{Constraining the LSBG's mass}\label{sec:modeling}

In this section, we present the models used to interpret the tangential shear measurements. For the red LSBG sample, we fit a model to the measurements to constrain the host halo and subhalo masses, as well as the offset between the host halo and subhalo. For the blue LSBG sample, we constrain only the host halo mass. We primarily focus on the analysis for the red LSBGs, due to their higher signal-to-noise.

 We note that when quantitatively analyzing the differences between models, we use the $\Delta \chi^2$ metric, defined as $$(\gamma_{t_{m1}} - \gamma_{t_{m2}}) \cdot \mathrm{C}^{-1} \cdot (\gamma_{t_{m1}} - \gamma_{t_{m2}})^\mathrm{T},$$ where $\gamma_{t_{m1}}$ and $\gamma_{t_{m2}}$ represent two different shear models. A $\Delta \chi^2$ of 1 indicates a $1\sigma$ shift between models.

\subsection{Modeling the Tangential Shear}
\label{sec:model}
The tangential shear for a system with sources at redshift $z_S$ and lenses at redshift $z_L$ can be written as 
\begin{equation}\label{eq: shearmodel} 
\gamma_{t} (z_S, z_L, \theta ) = \frac{\Delta \Sigma (R(\theta, z_{L}))}{\Sigma_{\mathrm{crit}}(z_{L}, z_{S})}. 
\end{equation} 
Here, $\Delta \Sigma (R(\theta))$ is the excess surface density as a function of the projected radius $R$, the distance between the lens and the source galaxy at the lens redshift. We obtain $R$ by translating from the observed angular space ($\theta$) to the physical space using $R=\theta \times D_L$, where $D_{L}$ is the angular diameter distance to the lens redshift $z_L$. 

The excess surface density measures the amount of mass above the average surface mass density within a region and determines the deflection of light for gravitational lensing events. $\Sigma_\mathrm{crit}(z_L, z_S)$ describes a geometrical factor dependent on the characteristics of the lensing system, given by 
\begin{equation}\label{eq:sigmacrit} 
\Sigma_\mathrm{crit}(z_L, z_S) = \frac{c^2}{4\pi G}\frac{D_S}{D_{LS} D_L}, ~\mathrm{if}~ z_S>z_L. 
\end{equation} 
Here $c$ represents the speed of light, while $D_S$ and $D_{LS}$ are the angular diameter distances to the source and between the source and the lens. If $z_S<z_L$, $\Sigma_\mathrm{crit}(z_L, z_S) =0$. 

The excess surface density can be broken into two components: 
\begin{equation}\label{eq:excessprojecteddensity}
\Delta\Sigma(R) = \Sigma(<R) - \Sigma(R),
\end{equation}
where $\Sigma (R)$ designates the projected surface density at $R$ and $\Sigma (<R)$ represents the cumulative surface density within the projected radius $R$. Given their dependence on $R$, both terms implicitly rely on the lens redshift. We define the cumulative surface density, $\Sigma(<R)$, as \begin{equation}\label{eq:cumulativesurfacedensity}\Sigma(<R) = \frac{2}{R^2}\int_0^R \chi \Sigma(\chi)d\chi,\end{equation} where $\chi$ is the distance perpendicular to the line of sight.

In this work, we assume both the host halo and subhalo associated with the lens galaxies take the form of Navarro-Frank-White (NFW) profiles \citep{Navarro1996}, where the density profile can be described as 
\begin{equation} \label{eq:NFWprofile}
\rho _{\mathrm{NFW}}(r) = \frac{\rho_s}{r/r_s(1+r/r_s)^2}.
\end{equation} 
Here $r_s= \frac{r_{200}}{c_s}$ defines the scale radius, where $c_s$ is the concentration of the halo and $r_{200}$ is the virial radius. $\rho_s$ represents the characteristic density of the halo and $r$ is the distance from the center of the halo. We note that $\rho_{s}$ is related to the  halo mass $M_\mathrm{200c}$, or the halo mass at 200 times the critical density of the universe, via
\begin{equation}\label{eq:characteristicdensity} 
\rho_s = \frac{M_\mathrm{200c}}{4\pi r_s^3(\ln(1+c_s)-c_s/(1+c_s))}.
\end{equation} 
We define the concentration by applying the \citet{Ishiyama2021} concentration-mass relationship. We select this relationship due to its flexibility and applicability to low-redshift and low-mass galaxies. \footnote{ We test additional concentration-mass relationships, from \citet{Diemer2019} and \citet{Ludlow2016}. We find a $\Delta \chi^2$ of $1.49\times10^{-5}$ between the fiducial model and the \citet{Diemer2019} model and 0.004 between the fiducial model and the \citet{Ludlow2016} model. In addition, we find that the concentration is poorly constrained when included as a free parameter (see Sec.~\ref{sec:fitting the model}), but the results remain consistent with the fiducial model.} The projected surface density of an NFW halo centered at the origin can then be written as 
\begin{equation}\label{eq:projectedsurfacedensity}
 \Sigma(R) = 2\int_0^{+\infty} \rho_{\mathrm{NFW}}(\sqrt{s^2+R^2})ds,
\end{equation} 
where $s$ is the distance along the line of sight.
If the NFW halo is offset from the origin, we have instead
\begin{equation}\label{offsetprofile} 
 \Sigma_\mathrm{off}(R, R_\mathrm{off}) = \frac{1}{2\pi}\int_0^{2\pi} \Sigma\left(\sqrt{R^2_\mathrm{off} +R^2 +2RR_\mathrm{off}\cos\theta}\right)d\theta, 
\end{equation}
where $R_\mathrm{off}$ is the projected distance between the new reference and the center of the profile. $\theta$ is the angle between the offset direction and the radial direction of $R$ where the line-of-sight projection is taken. In practice, we translate $R_{\rm off}$ to an offset angular scale $\theta_{\rm off}$ via $\theta_{\rm off} = R_{\rm off}/ D_L$. 

To account for the ensemble of source and lens galaxies, we integrate the tangential shear model over the redshift distributions to construct a model with a range of possible source and lens redshifts: 
\begin{equation}\label{eq:shearintegral} 
\gamma_{t,\mathrm{pop}}(\theta) = \int \int   \gamma_{t}(z_S, z_L, \theta)  n(z_S)  n(z_L) dz_S dz_L, 
\end{equation} 
where $n(z_L)$ and $n(z_S)$ represent the normalized redshift distributions for the lenses and sources. 

 We use the lens redshift distribution derived in Sec.~\ref{sec:cross correlation measurements}. We resample the redshift distribution over its jackknife covariance and find that incorporating these resampled redshift distributions into the shear model produces a small effect, inducing a $\Delta \chi^2 < 0.08$ 68\% of the time. In the fiducial model, we fix the lens redshift distribution for simplicity. Regarding the sources, we find that using a single fixed source redshift versus the full distribution of source redshifts does not significantly impact the model, resulting in a $\Delta \chi^2$ of 0.10. To reduce computational time, we fix the source redshift at the mean redshift of $z =0.6312$. We further verify that the uncertainty for the mean source redshift is negligible in Appendix~\ref{sec:testingpriors}. We find that combining these uncertainties (i.e. resampling over the lens redshift distribution and using the full source redshift distribution) produces a $\Delta \chi^2<0.17$ 68\% of the time.

As discussed at the end of Sec.~\ref{sec:measurements} and shown in Appendix \ref{sec:hodmodels}, there are compelling reasons to believe that the red LSBG sample is predominantly composed of satellite galaxies. This suggests that to first order, the total tangential shear is composed of two components, one representing the LSBG subhalo, $\gamma_\mathrm{t, pop}^\mathrm{sub}(\theta)$, with excess surface density $\Delta \Sigma^\mathrm{sub}(R)$, and one representing the host halo, $\gamma_\mathrm{t,pop}^\mathrm{host}(\theta, \theta_\mathrm{off})$, with excess surface density $\Delta \Sigma_\mathrm{off}^\mathrm{host}(R, R_\mathrm{off})$. See \citet{Li2013}, \citet{Sifon2018}, and \citet{wang2024} for other examples of modeling the galaxy-galaxy lensing signal around satellite galaxies.  We sum the subhalo and host halo terms to define the total tangential shear model: \begin{equation}\label{eq:totalshearmodel} \gamma_t^\mathrm{tot} (\theta, \theta_{\rm off}) = \gamma_\mathrm{t,pop}^\mathrm{sub}(\theta) + \gamma_\mathrm{t,pop}^\mathrm{host}(\theta, \theta_{\rm off}).\end{equation}

Finally, to represent the host halos of a galaxy population with various offset radii, we assume that the distribution of the offset in angular space is a normal distribution centered around $\theta_{\rm off}$ with a spread of $\theta_{\rm off}/3$. This parametrization appears to be a good description of the data (see Appendix~\ref{sec:testingpriors} and Fig.~\ref{fig:mcmc_offsetwidth} for further validation tests). 

We implement the above model using \textsc{Profiley} \citep{Madhavacheril2020}, a Python-based package that generates mass distribution profiles for galaxies. 

\subsection{Parameter Inference}\label{sec:fitting the model}

\begin{table*}
\centering
\begin{tabular}{ccccc}
\hline
Parameter & Priors & Posteriors & $\chi^2_\mathrm{best}/\nu$ & $\chi^2_\mathrm{upper}/\nu$ \\
\hline
 & Red LSBG & & $10.60/22$ & $26.16/22$ \\
	$M_{\rm sub}$ & U($7, 12$)  & $\log(M_\mathrm{sub}/M_\odot)<11.51$ \\
    $M_{\rm host}$ & U($10, 15$)  & $\log(M_\mathrm{host}/M_\odot)=12.98^{+0.10}_{-0.11}$ \\ 
   $\theta_{\rm off}$ & U(25,55)  & $36.6^{+4.7}_{-4.2}~\mathrm{arcmin}$ \\
\hline
& Blue LSBG & & $4.10/15$ & $24.97/15$ \\
$M_{\rm host}$ & U($10, 15$)  & $\log(M_\mathrm{host}/M_\odot)<11.84$ \\ 
\hline
\end{tabular}
\caption{\textit{Top three rows:} Priors and mean posteriors for the red LSBG tangential shear measurements for the subhalo mass, host halo mass, and radial offset given by the MCMC, corresponding to Fig.~\ref{fig:mcmc}. $\chi^2_\mathrm{best}/\nu$ and $\chi^2_\mathrm{upper}/\nu$ indicate the $\chi^2$ of the models with the respective best fit and upper bound parameters, divided by the number of degrees of freedom. \textit{Bottom row:} Prior and posterior for the blue LSBG tangential shear measurements for the host halo mass, shown in Fig.~\ref{fig:red_blue_galaxy_shear_model}. Note that both the red subhalo mass posterior and the blue host halo mass posteriors are unconstrained at the low mass ends, thus we list the 95\% upper bound.  We define $U(a,b)$ as a uniform distribution with a lower bound of $a$ and an upper bound of $b$.} 
\label{tab:model_params}
\end{table*}

We fit the model above to the LSBG tangential shear measurements using a Markov Chain Monte Carlo (MCMC) approach. 
We assume a Gaussian likelihood $L$, with 
\begin{equation} \label{eq:likelihood} 
\ln ( L ( \gamma_{td} | \boldsymbol{p}) )
= -\frac{1}{2}(\gamma_{td} - \gamma_{tm}(\textbf{p}))\textbf{C}^{-1} (\gamma _{td} - \gamma_{tm}(\textbf{p})), 
\end{equation} 
where $\gamma_{td}$, $\gamma_{tm}$ and $\textbf{C}^{-1}$ are the same as that defined in Eq.~\ref{eq:chi2}. 
The Bayesian posterior is proportional to the likelihood times the prior, or 
\begin{equation} 
P(\textbf{p}|\gamma_{td}) \propto L ( \gamma_{td} | \boldsymbol{p})P(\textbf{p}), 
\end{equation} 
where $P(\textbf{p})$ represents the prior. For the red LSBG sample, we have three free parameters in this model, \begin{equation}\label{eq:priors} 
\textbf{p} = (\log(M_\mathrm{sub}/M_\odot), \log(M_\mathrm{host}/M_\odot), \theta_\mathrm{off}), 
\end{equation} where $\log(M_\mathrm{sub}/M_\odot)$ and $\log(M_\mathrm{host}/M_\odot)$ follow the mass definition of Eq.~\ref{eq:characteristicdensity}. The priors on these parameters are listed in Table~\ref{tab:model_params}. For the MCMC, we use the \textsc{emcee} package with 20 walkers with a burn-in of 5000 steps and a chain length of 200,000 steps \citep{Goodman2010,Foreman-Mackey2013}.

We use \textsc{ChainConsumer} \citep{Hinton2016} to process the output from the MCMC.
We show the posterior distribution of the MCMC in Fig.~\ref{fig:mcmc}. We find the best-fit model corresponds to a mean offset of $\theta_{\rm off} = 36.6 ^{+4.7}_{-4.2}$ arcminutes and a host halo mass $\log(M_{\rm host}/M_{\odot}) = 12.98 ^{+0.10}_{-0.11}$. Based on the $n(z)$ distribution of Sec.~\ref{sec:lens catalog}, this mean offset corresponds to a physical distance range of $[0.23^{+0.03}_{-0.03}\mathrm{Mpc}$,$5.42^{+0.70}_{-0.62}\mathrm{Mpc}]$. Our data does not fully constrain the subhalo mass parameter, consistently pushing the best fit to the bottom edge of the prior space at $\log(M_\mathrm{sub}/M_\odot) = 7$. Since we cannot successfully constrain the mass at this lower bound, to first approximation this bound is irrelevant. The subhalo and host halo mass priors cross several orders of magnitude, requiring us to sample in a logarithmic space. We note that constraining a partially unbounded parameter in a flat logarithmic space can artificially lower the upper bound \citep{Diacoumis2019}. To avoid this problem and successfully obtain the upper limit corresponding to a 95\% confidence interval for the subhalo mass, we sample directly from the likelihood. We divide the subhalo parameter space into 100 thin slices and use the optimizer \texttt{iminuit} \citep{James1975} to find the host halo mass and radial offset parameters corresponding to the maximum likelihood for each subhalo mass slice. We obtain a subhalo mass upper bound of $\log(M_{\rm sub}/M_{\odot}) \leq 11.51$, an associated host halo mass of  $\log(M_{\rm host}/M_{\odot}) = 12.96$, and a mean radial offset of 37.14 arcmin. We find a $\chi^2$ of $10.60/22$ for the best-fit model and a $\chi^2$ of $26.16/22$ for the upper-bound model. The best-fit model yields a low $\chi^2/\nu$ value with a p-value of 0.98. This result suggests a potential overestimation of uncertainties or biases derived from inverting the noisy jackknife covariance, assuming the low $\chi^2$ value is not due to a statistical fluctuation, which cannot be entirely ruled out.

\begin{figure}
    \centering
    \includegraphics[width=0.43\textwidth]{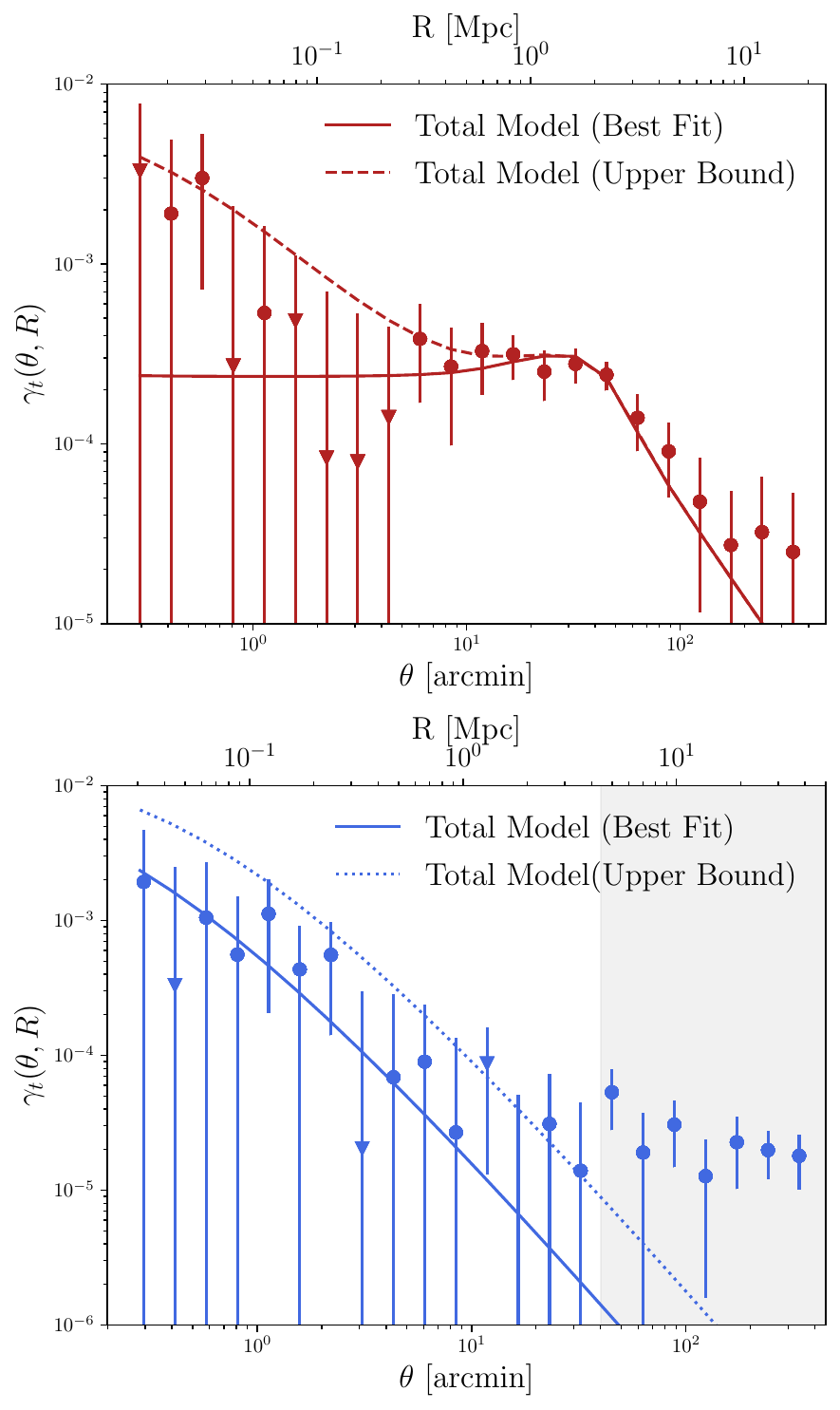}
    \caption{\textbf{Top}: Total model fit to the red LSBG tangential shear measurements. We compare the model produced with the best-fit parameters (solid line) to the model produced with the subhalo mass upper bound and associated host-halo mass and radial offset parameters (dashed line), as listed in Table~\ref{tab:model_params} and described in Sec.~\ref{sec:fitting the model}. Note that the triangle markers indicate a negative shear measurement. \\
    \textbf{Bottom}: 
    Total model fit to the blue galaxy tangential shear measurements. Angular scales above $\sim 50$ arcminutes are shaded in gray to indicate that these larger scales reach the two-halo term at the derived $n(z_L)$. We show the model for visualization purposes, but note that the two-halo term is not accounted for in our modeling. With a halo mass of $\log(M_\mathrm{host}/M_\odot)<11.84$, we find a $\chi^2$ value of 24.97 over 15 degrees of freedom between the model and the measurements. The solid line represents the model produced with the best-fit host halo mass, and the dashed line signifies the model produced with the upper-bound host halo mass.}
    \label{fig:red_blue_galaxy_shear_model}
\end{figure}

\begin{figure}
    \centering
    \includegraphics[width=0.45\textwidth]{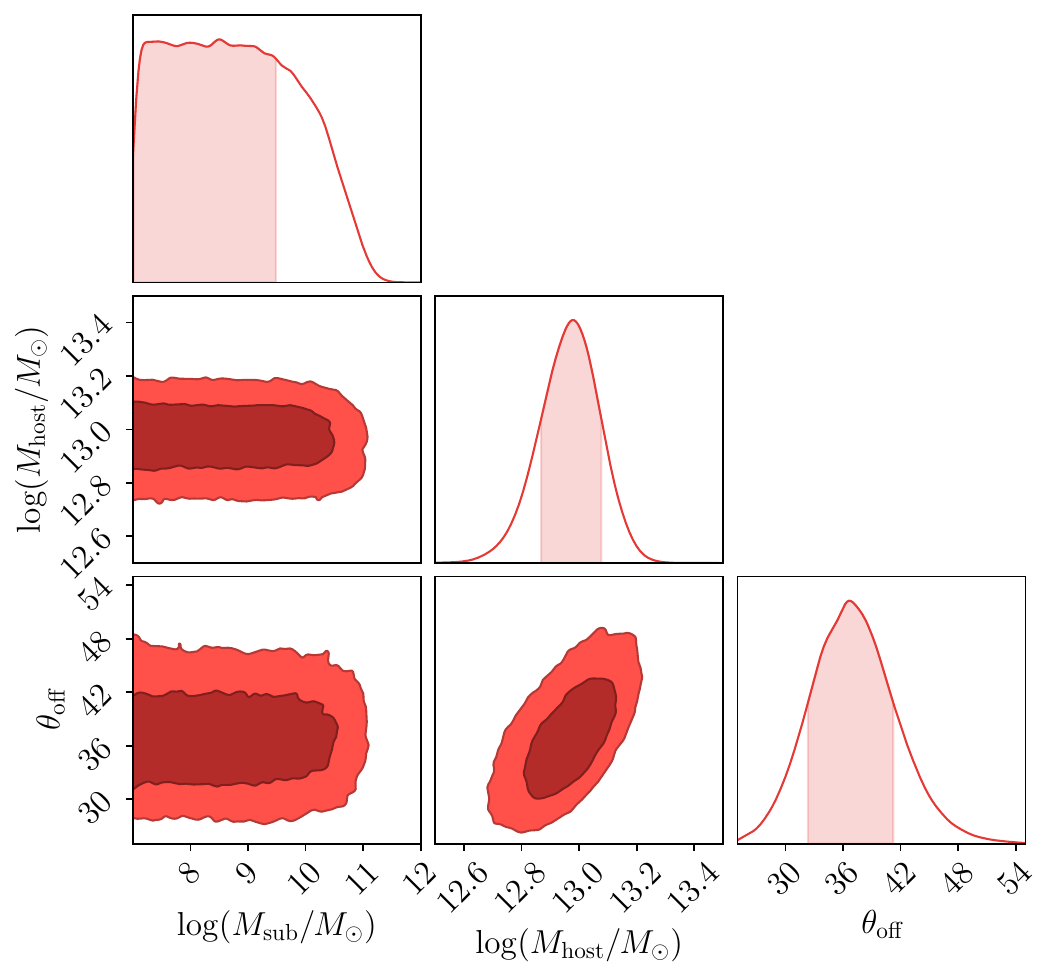}
    \caption{MCMC posterior distribution for red LSBGs. The free parameters include the subhalo mass, $M_\mathrm{sub}$, the host halo mass, $M_\mathrm{host}$, and the mean of the radial offset distribution, $\theta_\mathrm{off}$. The host halo mass is constrained at $\log(M_\mathrm{host}/M_\odot) = 12.98^{+0.10}_{-0.11}$ and the mean radial offset at $36.6^{+4.7}_{-4.2}~\mathrm{arcmin}$. The subhalo mass is unconstrained at the lower bound.}
    \label{fig:mcmc}
\end{figure}

As described in Sec.~\ref{sec:measurements}, we assume the blue LSBG sample is dominated by central galaxies, thus we utilize a single NFW profile to model the tangential shear measurements and vary only the host halo mass. We incorporate the derived blue LSBG redshift distribution shown in Fig.~\ref{fig:redshift_distribution_avg} into the model. Based on this higher redshift distribution, we remove scales greater than 50 arcmin (4.2 Mpc at $z=0.1$) from the blue LSBG shear measurements to avoid the two-halo term regime. We use the same host halo prior as the red LSBGs, $U(10,15)$. We tested a shifted prior of $U(8,14)$, closer to the red LSBG subhalo mass prior, but found that the results remained consistent. Similarly to the red LSBG subhalo term, the data does not fully constrain the blue LSBG host halo mass. We obtain an upper mass limit with 95\% certainty at a mass of $\log(M_\mathrm{host}/M_\odot) <11.84$, corresponding to a $\chi^2$ of 24.97/15. We find a $\chi^2$ of $4.10/15$ at the best-fit halo mass of $\log(M_\mathrm{host}/M_\odot) = 11$. We present the model of the blue LSBG sample in Fig.~\ref{fig:red_blue_galaxy_shear_model}.

\subsection{Discussion of the Results}
\label{sec:result_discussion}
When comparing our results to recent literature, we find that the red subhalo mass upper bound concurs with the \citet{Sifon2021} upper bound of $\log(M_{\rm sub}/M_{\odot})\leq 11.80$.
Though the \citet{Sifon2021} sample specifically refers to ultra-diffuse galaxies (UDGS), defined as LSBGs with a physical radius larger than 1.5~kpc and a central surface brightness larger than $24~\mathrm{mag}~\arcsec^{-2}$, the coinciding results indicate that our mass bound is reasonable. We estimate around 25\% (15\%) of the red (blue) LSBGs in our sample are UDGs (see Appendix~\ref{sec:UDGs} for more details). 
In addition, \citet{vanDokkum2016} estimated the subhalo mass of the UDG Dragonfly 44 at $8 \times 10^{11} M_{\odot}$, or $\log(M/M_{\odot})=11.90$. Considering Dragonfly 44's classification as a massive UDG, our results fall well within this range. Finally, our subhalo mass bound resides within the span described in \citet{Prole2019}, which presented a range of UDG and LSBG subhalo masses derived using the number counts of globular clusters. Our LSBG subhalo mass bound thus appears well aligned with previous studies of LSBG and UDG halo masses.

Similarly, we can compare our host halo mass upper bound for the blue LSBG sample to both isolated dwarf galaxy masses and the general central galaxy population. \citet{Thornton2023} used weak lensing to extract halo mass profiles from a sample of low-redshift dwarf galaxies in DES. They found that the median of their halo mass varied between $\log(M_\mathrm{host}/M_\odot) = [10.67^{+0.2}_{-0.4}, 11.40^{+0.08}_{-0.15}]$. The upper bound of this range fits between the best fit and upper bound estimates of the blue LSBG halo masses. \citet{Mandelbaum2015} reviewed the estimates of host halo masses derived from weak lensing for several works and found a range corresponding to $\log(M_\mathrm{vir}/M_\odot) = [11.0, 13.5]$. We note that, given the low signal-to-noise, our upper bound for the blue LSBG host halo mass is not particularly stringent. Nonetheless, we are generally compatible with current literature and future work will produce higher signal-to-noise shear measurements, enabling more precise results. 

\section{Stellar-to-Halo-Mass Relation}\label{sec:SHMR}

The stellar-to-halo-mass relation (SHMR) describes the connection between the mass of dark matter halos and their resident galaxies \citep{Moster2010, Behroozi2010}. The SHMR can be used to study the growth of dark matter halos, constrain theoretical models of galaxy formation, and gain insight into cosmological relationships between dark matter halos and large scale structures \citep{Girelli2020}. Comparing the SHMR of LSBGs to the SHMR of the wider galaxy population may reveal whether LSBGs have fundamentally unique characteristics \citep{Moster2010, Niemiec2022}. To evaluate the SHMR of the LSBGs, we first estimate their stellar mass in Sec.~\ref{sec:stellar mass}. Combining the stellar mass estimates with the halo mass results from Sec.~\ref{sec:modeling}, we present SHMR results in Sec.~\ref{sec:subSHMR}.

\subsection{Stellar Mass}\label{sec:stellar mass}

\begin{figure}
    \centering
    \includegraphics[width=0.45\textwidth]{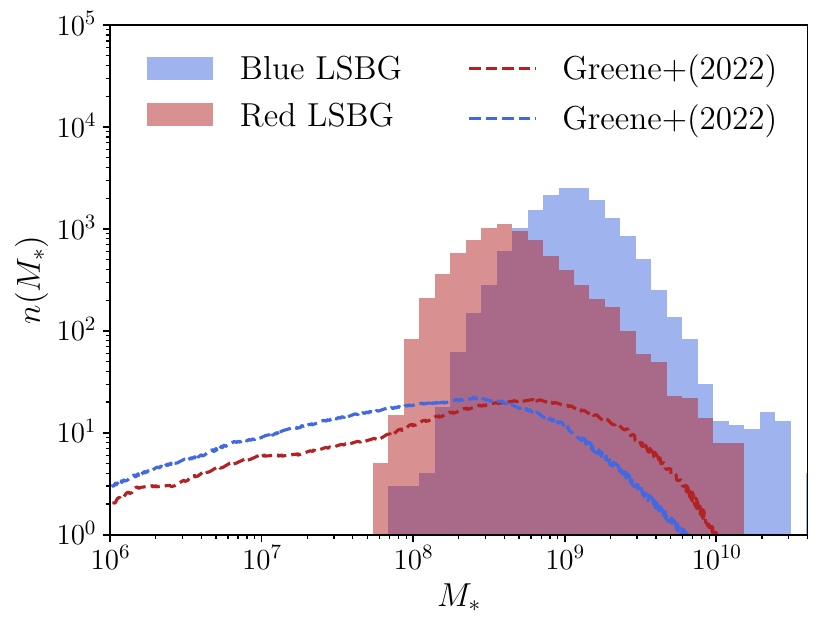}
    \caption{Comparison between the red and blue LSBG stellar mass distributions, generated with the \citet{Bell2003} CMLR, and the results from the HSC LSBG red and blue stellar mass distributions from \citet{Greene2022}, represented by the dashed lines. Note that the \citet{Greene2022} distribution has a larger low-mass tail, possibly due to the increased depth of the HSC data, and utilizes the \citet{Into2013} CMLR.}
    \label{fig:stellar_mass_distribution}
\end{figure}

To estimate the stellar mass of the red LSBG sample, we adopt the method developed in \citet{Bell2003}. We use the color-mass-to-light relation (CMLR) to obtain stellar mass estimates from apparent magnitudes. This relation is usually parameterized as follows:
\begin{equation} \label{eq:CMLR} 
\log \gamma_* ^j = a_j + b_j \times \mathrm{color}. 
\end{equation} 
Here, $\gamma_*^j$ represents the stellar mass-to-light ratio, $j$ stands for the selected band, $a_j$ indicates the zero-point of the function, and $b_j$ signifies the slope of the function. 

An extensive suite of CMLR relations can be found in the literature.  \citet{Du2020} re-calibrated such a relation for a sample of LSBGs selected from SDSS \citep{Du2019} by fitting their SEDs (spectral energy distributions) to a stellar population synthesis (SPS) model. They concluded that their LSBG population is likely to follow a representative CMLR relation defined on diverse galaxy populations. Their sample has a higher surface brightness cutoff (with $\mu > 22.5$), but they used a similar color distribution to our population. Following these conclusions, we proceed to use the parameter values from the standard \citet{Bell2003} relation listed in Table ~\ref{tab:CMLR_parameters}. 

Next, we calculate the LSBG's luminosity $L$ for a given band $j$ from that band's apparent magnitude. 
Then, we convert the redshift, $z$, for each point in the $n(z)$ to a comoving distance, $d$, and an absolute \textit{j}-band magnitude for each galaxy via
\begin{equation} \label{eq:distancemodulus} 
M_j (m_j, z_L) = m_j - 5 \log ( \frac{d(z_L)}{10 \mathrm{pc}}), 
\end{equation}
which we integrate over the redshift distribution to obtain the absolute magnitude distribution, 
\begin{equation} \label{eq:weighredmagnitude}
M_j (m_j)=\int{M_j (m_j, z_L) n(z_L) dz_L}. 
\end{equation}

Next, we convert the absolute magnitude distribution to a luminosity distribution in a given band $j$ via 
\begin{equation} \label{eq:luminosity} 
\frac{L_j}{L_{j, \odot}} = 10^{0.4(M_{j, \odot} - M_j}), 
\end{equation}
where $M_{j, \odot}$ is the absolute magnitude of the Sun for the AB magnitude SDSS filter in the given band $j$. Finally, the stellar mass distribution of the LSBG sample can be derived via 
\begin{equation}\label{eq:stellarmassestimate} 
{M_*} = L_j \times \gamma_*^j. 
\end{equation} 
We carried out this procedure for the $g$, $r$, and $i$ magnitudes and for the $g-i$ and $g-r$ colors and found that the the stellar mass distribution estimate remained consistent regardless of the selected colors or bands. Therefore, we follow \citet{Du2020} and use the $g-r$ color and $i$-band for our analysis.

With this process, we find the stellar mass distributions shown in Fig.~\ref{fig:stellar_mass_distribution}. The median stellar mass is $4.2\times 10^{8} M_\odot$ and the mean is $6.8\times 10^8 M_\odot$ for the red LSBG sample. For the blue LSBG sample, the median stellar mass is $1.11\times 10^9 M_\odot$ and the mean is $1.46 \times 10^9 M_\odot$.

\subsubsection{Comparison to Previous Work} 
\label{sec:stellar_mass_previous_comparison}
\citet{Greene2022} utilized color to estimate stellar masses for a comparable LSBG sample identified in the HSC survey. They used the CMLR from \citet{Into2013}, as opposed to the \citet{Bell2003} relation, but explained that they are currently computing a new color-mass relation for a set of H I selected UDGs and NASA-Sloan Atlas \citep[NSA,][]{Blanton2005} dwarfs. They estimated this relation to be between the \citet{Into2013} and \citet{Bell2003} results. We compared the stellar masses generated using both the \citet{Into2013} and \citet{Bell2003} CMLRs and found that the distributions remained consistent. We compare our stellar mass distributions with the results from  \citet{Greene2022} in Fig.~\ref{fig:stellar_mass_distribution}.

\citet{Sifon2021} also estimated the median stellar mass of their UDG sample at $2\times 10^8 M_\odot$, similar to the median of our red LSBG stellar mass distribution. The SAGA survey \citep{Mao2021} of satellite dwarf galaxies presented a median stellar mass of $9.5 \times 10^7 M_\odot$. Given SAGA's lower redshifts and nature as a dwarf-specific survey, the lower median stellar mass is sensible, though still contained within the span of our stellar mass range. The dwarf-focused \citet{Thornton2023} sample provided median stellar masses of $[10^{8.52}M_\odot, 10^{9.02}M_\odot, 10^{9.49}M_\odot]$, corresponding to low, middle, and high-mass bins, that align with our estimate.

\begin{table}
\centering
\begin{tabular}{ c c c c c c c} 
 \hline
  Color & $a_g$ & $b_g$ & $a_r$ & $b_r$ & $a_i$ & $b_i$ \\ 
 \hline
 $g-r$  & -0.499 & 1.519 & -0.306 & 1.097 & \textbf{-0.222} & \textbf{0.864} \\ 
 \hline
 $g-i$ & -0.379 & 0.914 & -0.220 & 0.661 & -0.152 & 0.518 \\
 \hline
 \bottomrule
\end{tabular}
\caption{Zero-point and slope parameters for the CMLR \citep{Bell2003} for the \textit{g}-band, \textit{i}-band, and \textit{r}-band magnitudes and the $g-r$ and $g-i$ colors. We utilize the bolded values in our work.}
\label{tab:CMLR_parameters}
\end{table}

\subsection{Stellar to Halo Mass Relation for Red and Blue LSBG Samples} \label{sec:subSHMR} 

\begin{figure}
    \centering
    \includegraphics[width=0.45\textwidth]{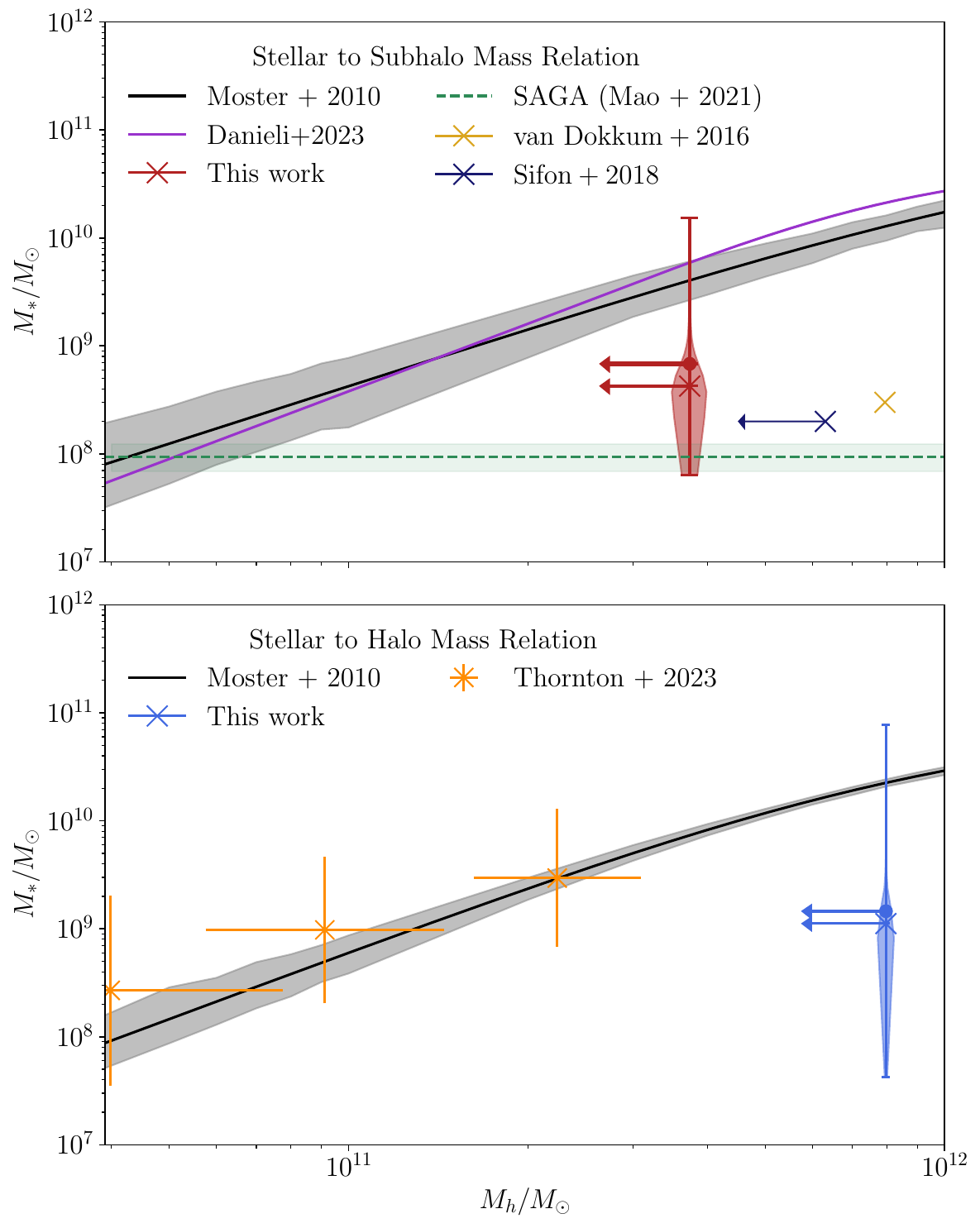}
    \caption{
    Constraints on the stellar-to-subhalo mass relation (red violin, \textit{top}) and stellar-to-halo mass relation (blue violin, \textit{bottom}) for the red and blue LSBG samples, respectively. The cross marking represents the median and the circle marking indicates the mean of the LSBG sample stellar mass distribution. The horizontal red and blue arrows signify that the violins are placed at the upper limits on the halo mass. We include comparisons with the parameterized SHMR by \citet{Moster2010} (black line) and, in the case of the red sample, \citet{Danieli2023} (purple line). The grey shading represents the uncertainty of the parameters for \citet{Moster2010}. The dark blue cross marks the median of the \citet{Sifon2018} stellar mass distribution, positioned at the upper bound of the subhalo virial mass found in that work. The dashed green line stands for the median of the SAGA \citep{Mao2021} stellar mass distribution, for which we have no subhalo mass information. The gold cross shows the stellar-to-halo mass relation for the UDG Dragonfly 44 galaxy \citep{vanDokkum2016} and the orange crosses show the stellar-to-halo mass relation for the \citet{Thornton2023} dwarf galaxy sample. We note that $\sim 20\%$ of the \citet{Thornton2023} sample is composed of satellite galaxies.}
    \label{fig:SHMR}
\end{figure}

Provided with stellar mass estimates, we can now calculate the stellar-to-halo mass relation for the LSBG samples. To find the SHMR, we divide $M_*/M_{h, vir},$ where $M_{h, vir}$ represents the virial halo mass and $M_*$ indicates the stellar mass. Since the red LSBG sample is dominated by satellite galaxies, we obtain constraints for the subhalo-to-halo mass relation. For the central-dominated blue LSBG sample, $M_h$ represents the host halo mass, $M_\mathrm{host}/M_\odot$. We show the median, mean, and spread of the stellar mass distribution for the red and blue LSBG samples at their respective upper subhalo and host halo mass bounds, converted to virial masses of $3.7 \times 10^{11} M_\odot$ and $7.9\times 10^{11} M_\odot$, in Fig.~\ref{fig:SHMR} \footnote{We note that we technically measure $P(M_h|M_*)$, or the halo mass for a sample with a given stellar mass.}. We compare our results with the measurements from \citet{vanDokkum2016}, \citet{Sifon2021}, \citet{Thornton2023}, and the stellar mass estimate of \citet{Mao2021}, as described in Sec.~\ref{sec:result_discussion} and Sec.~\ref{sec:stellar_mass_previous_comparison}, due to similarities in their target galaxy sample characteristics.

We can also compare our results to more general, parameterized SHMRs. We contrast the red LSBG SHMR against the \citet{Danieli2023} SHMR and the \citet{Moster2010} satellite-specific SHMR. \citet{Danieli2023} used the ELVES satellite galaxy survey \citep{Carlsten2022} and the semianalytical \texttt{SatGen} model \citep{Jiang2021} to present constraints on the connection between satellite galaxies and their respective dark matter subhalos. \citet{Moster2010} statistically determined the relationship between the stellar masses of galaxies and their resident dark matter halos by populating halos and subhalos in an N-body simulation with galaxies and reproducing the observed stellar mass function. We find that our constraint of the SHMR for the red sample is consistent with these more general SHMRs.  We uncover a similar result when comparing the blue LSBG SHMR to the well-established \citet{Moster2010} central-galaxy SHMR. We note that, given the constraining power of our sample, we are only able to distinguish the formation channels of LSBG samples from those of the general galaxy population if they exhibit significant divergence.

\section{Conclusions and Outlook} \label{sec:conclusion}
We perform galaxy-galaxy lensing measurements on a sample of LSBGs discovered in DES by \citet{Tanoglidis2021} and fit a simple model to these measurements to glean information about the mass of the LSBGs. 
We use the DES Y3 \metacal\ catalog for the source galaxy shapes and divide the lens sample into red ($g-i\ge0.6$) and blue ($g-i<0.6$) subsamples. We cross-correlate the positions of the red and blue LSBGs with the 2MPZ catalog to estimate the redshift distribution of the samples. We measure the tangential shear around the lens galaxies across angular scales of 0.25-400 arcmin and extract signal-to-noise ratios of 6.67 for the red subsample, 2.17 for the blue subsample, and 5.30 for the combined sample.  

 We assume that the red LSBG sample is primarily composed of satellite galaxies and construct a model built out of two NFW profiles to represent the subhalo and host halo. We fit the model to the red LSBG shear measurements to recover posterior values for the host halo mass, the mean radial offset of the host halo for the satellite population, and the subhalo mass. We cannot fully constrain the subhalo mass, but we obtain an upper bound with 95\% certainty at $\log(M_\mathrm{sub}/M_{\odot})<11.51$. We estimate the host halo mass at $\log(M_\mathrm{host}/M_\odot) = 12.98^{+0.10}_{-0.11}$ and the offset at $36.6^{+4.7}_{-4.2} \, \mathrm{arcmin}$. These results are presented in Table ~\ref{tab:model_params}. The host halo mass posterior and the upper bound of the subhalo mass posterior are consistent with the results of current literature.

Conversely, we assume the blue LSBG sample is dominated by central galaxies. As such, our model utilizes a single NFW profile to represent the host halo and requires one free parameter, the halo mass. Due to the large uncertainty, we can only obtain an upper bound on the host halo mass with 95\% certainty  at $\log(M_\mathrm{host}/M_{\odot})<11.84$. 

We use the lens photometry to estimate the stellar mass of the red and blue LSBG samples by adopting the color-mass light relation of \citet{Bell2003}. We combine the stellar mass estimate with the subhalo mass posterior to obtain the red LSBG stellar-to-subhalo mass ratio. We repeat this process with the blue LSBG sample. We compare these measurements to more general, parameterized satellite and central stellar-to-halo mass relations. Given our uncertainties, we find that the red and blue LSBG sample's SHMRs are consistent with the  general SHMR.

This project represents the first attempt to constrain the mass range of LSBGs using weak gravitational lensing. These results present a possible litmus test for future lensing measurements conducted with upcoming projects like the Merian survey \citep{Luo2023}, the Euclid Wide survey \citep{Euclid2022}, the Roman survey \citep{McEnery2021}, or other deep and wide photometric surveys \citep{Leauthaud2020}.

Looking ahead, the DES Year 6 dataset will produce deeper, more expansive catalogs of both LSBGs and source galaxies. The source catalog will contain approximately 150M source galaxies with a number density of $8 \mathrm{gal}/\mathrm{arcmin}^2$, compared to 100M source galaxies and a number density of $5.59 \mathrm{gal}/\mathrm{arcmin}^2$ for DES Y3. The DES Y6 LSBG catalog may include upwards of 45,000 galaxies. With this increase in galaxy count, the lensing signal could grow by a factor of $\sim2.6$, improving the precision of the  LSBG mass constraints. The Dark Energy Spectroscopic Instrument \citep[DESI,][]{DESI2023} has already begun observation and will produce additional samples of LSBGs and dwarf galaxies. The Vera C.\ Rubin Observatory's Legacy Survey of Space and Time \citep[LSST,][]{Ivezic2019} will begin its survey in 2025, covering more than three times the area of DES and extending two magnitudes deeper. This work builds a foundation for similar future analyses with larger, deeper, and more constraining data sets.

\section*{Acknowledgments}

This work was supported in part by the U.S. Department of Energy (DOE), Office of Science, Office of Workforce Development for Teachers and Scientists (WDTS) under the Science Undergraduate Laboratory Internships Program (SULI).
JP has been supported by the Eric and Wendy Schmidt AI in Science Postdoctoral Fellowship, a Schmidt Futures program. CC is supported by NSF via award AST-2108168 and DOE via award DE-SC0021949. ADW is partially supported by NSF awards AST-2006340, AST-2108168, and AST-2307126.

Funding for the DES Projects has been provided by the U.S. Department of Energy, the U.S. National Science Foundation, the Ministry of Science and Education of Spain, 
the Science and Technology Facilities Council of the United Kingdom, the Higher Education Funding Council for England, the National Center for Supercomputing 
Applications at the University of Illinois at Urbana-Champaign, the Kavli Institute of Cosmological Physics at the University of Chicago, 
the Center for Cosmology and Astro-Particle Physics at the Ohio State University,
the Mitchell Institute for Fundamental Physics and Astronomy at Texas A\&M University, Financiadora de Estudos e Projetos, 
Funda{\c c}{\~a}o Carlos Chagas Filho de Amparo {\`a} Pesquisa do Estado do Rio de Janeiro, Conselho Nacional de Desenvolvimento Cient{\'i}fico e Tecnol{\'o}gico and 
the Minist{\'e}rio da Ci{\^e}ncia, Tecnologia e Inova{\c c}{\~a}o, the Deutsche Forschungsgemeinschaft and the Collaborating Institutions in the Dark Energy Survey. 

The Collaborating Institutions are Argonne National Laboratory, the University of California at Santa Cruz, the University of Cambridge, Centro de Investigaciones Energ{\'e}ticas, 
Medioambientales y Tecnol{\'o}gicas-Madrid, the University of Chicago, University College London, the DES-Brazil Consortium, the University of Edinburgh, 
the Eidgen{\"o}ssische Technische Hochschule (ETH) Z{\"u}rich, 
Fermi National Accelerator Laboratory, the University of Illinois at Urbana-Champaign, the Institut de Ci{\`e}ncies de l'Espai (IEEC/CSIC), 
the Institut de F{\'i}sica d'Altes Energies, Lawrence Berkeley National Laboratory, the Ludwig-Maximilians Universit{\"a}t M{\"u}nchen and the associated Excellence Cluster Universe, 
the University of Michigan, NSF's NOIRLab, the University of Nottingham, The Ohio State University, the University of Pennsylvania, the University of Portsmouth, 
SLAC National Accelerator Laboratory, Stanford University, the University of Sussex, Texas A\&M University, and the OzDES Membership Consortium.

Based in part on observations at Cerro Tololo Inter-American Observatory at NSF's NOIRLab (NOIRLab Prop. ID 2012B-0001; PI: J. Frieman), which is managed by the Association of Universities for Research in Astronomy (AURA) under a cooperative agreement with the National Science Foundation.

The DES data management system is supported by the National Science Foundation under Grant Numbers AST-1138766 and AST-1536171.
The DES participants from Spanish institutions are partially supported by MICINN under grants ESP2017-89838, PGC2018-094773, PGC2018-102021, SEV-2016-0588, SEV-2016-0597, and MDM-2015-0509, some of which include ERDF funds from the European Union. IFAE is partially funded by the CERCA program of the Generalitat de Catalunya.
Research leading to these results has received funding from the European Research
Council under the European Union's Seventh Framework Program (FP7/2007-2013) including ERC grant agreements 240672, 291329, and 306478.
We  acknowledge support from the Brazilian Instituto Nacional de Ci\^encia
e Tecnologia (INCT) do e-Universo (CNPq grant 465376/2014-2).

This manuscript has been authored by Fermi Research Alliance, LLC under Contract No. DE-AC02-07CH11359 with the U.S. Department of Energy, Office of Science, Office of High Energy Physics.

\subsection*{Author Contributions}
Nathalie Chicoine performed the main analysis, produced all the plots in the paper, and led the paper writing. Judit Prat provided direct supervision for the research and contributed to the paper’s text. Chihway Chang provided oversight for the supervision and contributed to the paper’s text. Georgios Zacharegkas generated the HOD model for the cross-code check and led the LSBG satellite composition analysis. Alex Drlica-Wagner contributed to the paper’s text and discussed problems relating to the LSBG sample properties, the stellar mass distribution, and the SHMR.  Dimitrios Tangolidis developed the LSBG catalog and assisted with the mask generation and development of the cross-correlation code. Dhayaa Anbajagane helped solve problems relating to the redshift distribution and the statistical analysis of the subhalo mass upper bound. Alexandra Amon, Susmita Adhikari, and Risa Weschler internally reviewed the paper.  The authors between A. Alarcon and J. Zuntz, both included, have produced and characterized one or multiple of the following data products used in this paper: the shape catalog, the redshift catalog, the Balrog image simulation, the PSF. Builders: The remaining
authors have made contributions to this paper that include, but are not limited to, the construction of DECam
and other aspects of collecting the data; data processing and calibration; developing broadly used methods,
codes, and simulations; running the pipelines and validation tests; and promoting the science analysis.

\bibliography{library}

\begin{thebibliography}{}
\makeatletter
\relax
\def\mn@urlcharsother{\let\do\@makeother \do\$\do\&\do\#\do\^\do\_\do\%\do\~}
\def\mn@doi{\begingroup\mn@urlcharsother \@ifnextchar [ {\mn@doi@}
  {\mn@doi@[]}}
\def\mn@doi@[#1]#2{\def\@tempa{#1}\ifx\@tempa\@empty \href
  {http://dx.doi.org/#2} {doi:#2}\else \href {http://dx.doi.org/#2} {#1}\fi
  \endgroup}
\def\mn@eprint#1#2{\mn@eprint@#1:#2::\@nil}
\def\mn@eprint@arXiv#1{\href {http://arxiv.org/abs/#1} {{\tt arXiv:#1}}}
\def\mn@eprint@dblp#1{\href {http://dblp.uni-trier.de/rec/bibtex/#1.xml}
  {dblp:#1}}
\def\mn@eprint@#1:#2:#3:#4\@nil{\def\@tempa {#1}\def\@tempb {#2}\def\@tempc
  {#3}\ifx \@tempc \@empty \let \@tempc \@tempb \let \@tempb \@tempa \fi \ifx
  \@tempb \@empty \def\@tempb {arXiv}\fi \@ifundefined
  {mn@eprint@\@tempb}{\@tempb:\@tempc}{\expandafter \expandafter \csname
  mn@eprint@\@tempb\endcsname \expandafter{\@tempc}}}

\bibitem[\protect\citeauthoryear{{Abbott} \& {Aguena} et~al.,}{{Abbott}
  et~al.}{2022}]{Abbott2022}
{Abbott} T.~M.~C.,  et~al. 2022, \mn@doi [\prd] {10.1103/PhysRevD.105.023520},
  \href {https://ui.adsabs.harvard.edu/abs/2022PhRvD.105b3520A} {105, 023520}

\bibitem[\protect\citeauthoryear{{Aihara} \& {AlSayyad} et~al.,}{{Aihara}
  et~al.}{2022}]{Aihara2022}
{Aihara} H.,  et~al. 2022, \mn@doi [\pasj] {10.1093/pasj/psab122}, \href
  {https://ui.adsabs.harvard.edu/abs/2022PASJ...74..247A} {74, 247}

\bibitem[\protect\citeauthoryear{{Bartelmann}}{{Bartelmann}}{2010}]{Bartelmann2010}
{Bartelmann} M.,  2010, \mn@doi [Classical and Quantum Gravity]
  {10.1088/0264-9381/27/23/233001}, \href
  {https://ui.adsabs.harvard.edu/abs/2010CQGra..27w3001B} {27, 233001}

\bibitem[\protect\citeauthoryear{{Bartelmann} \& {Maturi}}{{Bartelmann} \&
  {Maturi}}{2017}]{Bartelmann2017}
{Bartelmann} M.,  {Maturi} M.,  2017, \mn@doi [Scholarpedia]
  {10.4249/scholarpedia.32440}, \href
  {https://ui.adsabs.harvard.edu/abs/2017SchpJ..1232440B} {12, 32440}

\bibitem[\protect\citeauthoryear{{Behroozi}, {Conroy}  \&
  {Wechsler}}{{Behroozi} et~al.}{2010}]{Behroozi2010}
{Behroozi} P.~S.,  {Conroy} C.,   {Wechsler} R.~H.,  2010, \mn@doi [\apj]
  {10.1088/0004-637X/717/1/379}, \href
  {https://ui.adsabs.harvard.edu/abs/2010ApJ...717..379B} {717, 379}

\bibitem[\protect\citeauthoryear{{Bell} \& {McIntosh} et~al.,}{{Bell}
  et~al.}{2003}]{Bell2003}
{Bell} E.~F.,  et~al. 2003, \mn@doi [\apjs] {10.1086/378847}, \href
  {https://ui.adsabs.harvard.edu/abs/2003ApJS..149..289B} {149, 289}

\bibitem[\protect\citeauthoryear{{Berlind} \& {Blanton} et~al.,}{{Berlind}
  et~al.}{2005}]{Berlind2005}
{Berlind} A.~A.,  et~al. 2005, \mn@doi [\apj] {10.1086/431658}, \href
  {https://ui.adsabs.harvard.edu/abs/2005ApJ...629..625B} {629, 625}

\bibitem[\protect\citeauthoryear{{Bernstein} \& {Nichol} et~al.,}{{Bernstein}
  et~al.}{1995}]{Bernstein1995}
{Bernstein} G.~M.,  et~al. 1995, \mn@doi [\aj] {10.1086/117624}, \href
  {https://ui.adsabs.harvard.edu/abs/1995AJ....110.1507B} {110, 1507}

\bibitem[\protect\citeauthoryear{{Bhattacharyya} \& {Peter}
  et~al.,}{{Bhattacharyya} et~al.}{2023}]{Bhattacharyya2023}
{Bhattacharyya} J.,  et~al. 2023, \mn@doi [arXiv e-prints]
  {10.48550/arXiv.2312.00773}, \href
  {https://ui.adsabs.harvard.edu/abs/2023arXiv231200773B} {p. arXiv:2312.00773}

\bibitem[\protect\citeauthoryear{{Bilicki} \& {Jarrett} et~al.,}{{Bilicki}
  et~al.}{2014}]{Bilicki2014}
{Bilicki} M.,  et~al. 2014, \mn@doi [\apjs] {10.1088/0067-0049/210/1/9}, \href
  {https://ui.adsabs.harvard.edu/abs/2014ApJS..210....9B} {210, 9}

\bibitem[\protect\citeauthoryear{{Blanton} \& {Eisenstein} et~al.,}{{Blanton}
  et~al.}{2003}]{Blanton2003}
{Blanton} M.~R.,  et~al. 2003, in American Astronomical Society Meeting
  Abstracts. p. 145.01

\bibitem[\protect\citeauthoryear{{Blanton} \& {Lupton} et~al.,}{{Blanton}
  et~al.}{2005}]{Blanton2005}
{Blanton} M.~R.,  et~al. 2005, \mn@doi [\apj] {10.1086/431416}, \href
  {https://ui.adsabs.harvard.edu/abs/2005ApJ...631..208B} {631, 208}

\bibitem[\protect\citeauthoryear{{Bothun}, {Impey}  \& {McGaugh}}{{Bothun}
  et~al.}{1997}]{Bothun1997}
{Bothun} G.,  {Impey} C.,   {McGaugh} S.,  1997, \mn@doi [\pasp]
  {10.1086/133941}, \href
  {https://ui.adsabs.harvard.edu/abs/1997PASP..109..745B} {109, 745}

\bibitem[\protect\citeauthoryear{{Brook} \& {Di Cintio} et~al.,}{{Brook}
  et~al.}{2014}]{Brook2014}
{Brook} C.~B.,  et~al. 2014, \mn@doi [\apjl] {10.1088/2041-8205/784/1/L14},
  \href {https://ui.adsabs.harvard.edu/abs/2014ApJ...784L..14B} {784, L14}

\bibitem[\protect\citeauthoryear{{Carlsten} \& {Greene} et~al.,}{{Carlsten}
  et~al.}{2022}]{Carlsten2022}
{Carlsten} S.~G.,  et~al. 2022, \mn@doi [\apj] {10.3847/1538-4357/ac6fd7},
  \href {https://ui.adsabs.harvard.edu/abs/2022ApJ...933...47C} {933, 47}

\bibitem[\protect\citeauthoryear{{Cohen} \& {van Dokkum} et~al.,}{{Cohen}
  et~al.}{2018}]{Cohen2018}
{Cohen} Y.,  et~al. 2018, \mn@doi [\apj] {10.3847/1538-4357/aae7c8}, \href
  {https://ui.adsabs.harvard.edu/abs/2018ApJ...868...96C} {868, 96}

\bibitem[\protect\citeauthoryear{{Collister} \& {Lahav}}{{Collister} \&
  {Lahav}}{2004}]{Collister2004}
{Collister} A.~A.,  {Lahav} O.,  2004, \mn@doi [\pasp] {10.1086/383254}, \href
  {https://ui.adsabs.harvard.edu/abs/2004PASP..116..345C} {116, 345}

\bibitem[\protect\citeauthoryear{{DESI Collaboration} \& {Adame} et~al.,}{{DESI
  Collaboration}}{2024}]{DESI2023}
{DESI Collaboration} 2024, \mn@doi [\aj] {10.3847/1538-3881/ad3217}, \href
  {https://ui.adsabs.harvard.edu/abs/2024AJ....168...58D} {168, 58}

\bibitem[\protect\citeauthoryear{{Danieli} \& {van Dokkum} et~al.,}{{Danieli}
  et~al.}{2017}]{Danieli2017}
{Danieli} S.,  et~al. 2017, \mn@doi [\apj] {10.3847/1538-4357/aa615b}, \href
  {https://ui.adsabs.harvard.edu/abs/2017ApJ...837..136D} {837, 136}

\bibitem[\protect\citeauthoryear{{Danieli} \& {Greene} et~al.,}{{Danieli}
  et~al.}{2023}]{Danieli2023}
{Danieli} S.,  et~al. 2023, \mn@doi [\apj] {10.3847/1538-4357/acefbd}, \href
  {https://ui.adsabs.harvard.edu/abs/2023ApJ...956....6D} {956, 6}

\bibitem[\protect\citeauthoryear{{Dark Energy Survey Collaboration} \& {Abbott}
  et~al.,}{{Dark Energy Survey Collaboration}}{2016}]{DES2016}
{Dark Energy Survey Collaboration} 2016, \mn@doi [\mnras]
  {10.1093/mnras/stw641}, \href
  {https://ui.adsabs.harvard.edu/abs/2016MNRAS.460.1270D} {460, 1270}

\bibitem[\protect\citeauthoryear{{Das}}{{Das}}{2013}]{Das2013}
{Das} M.,  2013, \mn@doi [Journal of Astrophysics and Astronomy]
  {10.1007/s12036-013-9166-8}, \href
  {https://ui.adsabs.harvard.edu/abs/2013JApA...34...19D} {34, 19}

\bibitem[\protect\citeauthoryear{{Davis} \& {Peebles}}{{Davis} \&
  {Peebles}}{1983}]{Davis1983}
{Davis} M.,  {Peebles} P.~J.~E.,  1983, \mn@doi [\apj] {10.1086/160884}, \href
  {https://ui.adsabs.harvard.edu/abs/1983ApJ...267..465D} {267, 465}

\bibitem[\protect\citeauthoryear{{Dey} \& {Schlegel} et~al.,}{{Dey}
  et~al.}{2019}]{Dey2019}
{Dey} A.,  et~al. 2019, \mn@doi [\aj] {10.3847/1538-3881/ab089d}, \href
  {https://ui.adsabs.harvard.edu/abs/2019AJ....157..168D} {157, 168}

\bibitem[\protect\citeauthoryear{{Diacoumis} \& {Wong}}{{Diacoumis} \&
  {Wong}}{2019}]{Diacoumis2019}
{Diacoumis} J. A.~D.,  {Wong} Y. Y.~Y.,  2019, \mn@doi [\jcap]
  {10.1088/1475-7516/2019/05/025}, \href
  {https://ui.adsabs.harvard.edu/abs/2019JCAP...05..025D} {2019, 025}

\bibitem[\protect\citeauthoryear{{Diemer} \& {Joyce}}{{Diemer} \&
  {Joyce}}{2019}]{Diemer2019}
{Diemer} B.,  {Joyce} M.,  2019, \mn@doi [\apj] {10.3847/1538-4357/aafad6},
  \href {https://ui.adsabs.harvard.edu/abs/2019ApJ...871..168D} {871, 168}

\bibitem[\protect\citeauthoryear{{Driver}}{{Driver}}{1999}]{Driver1999}
{Driver} S.~P.,  1999, \mn@doi [\apjl] {10.1086/312379}, \href
  {https://ui.adsabs.harvard.edu/abs/1999ApJ...526L..69D} {526, L69}

\bibitem[\protect\citeauthoryear{{Du} \& {Cheng} et~al.,}{{Du}
  et~al.}{2019}]{Du2019}
{Du} W.,  et~al. 2019, \mn@doi [\mnras] {10.1093/mnras/sty2976}, \href
  {https://ui.adsabs.harvard.edu/abs/2019MNRAS.483.1754D} {483, 1754}

\bibitem[\protect\citeauthoryear{{Du} \& {Cheng} et~al.,}{{Du}
  et~al.}{2020}]{Du2020}
{Du} W.,  et~al. 2020, \mn@doi [\aj] {10.3847/1538-3881/ab6efb}, \href
  {https://ui.adsabs.harvard.edu/abs/2020AJ....159..138D} {159, 138}

\bibitem[\protect\citeauthoryear{{Euclid Collaboration} \& {Scaramella}
  et~al.,}{{Euclid Collaboration}}{2022}]{Euclid2022}
{Euclid Collaboration} 2022, \mn@doi [\aap] {10.1051/0004-6361/202141938},
  \href {https://ui.adsabs.harvard.edu/abs/2022A&A...662A.112E} {662, A112}

\bibitem[\protect\citeauthoryear{{Flaugher} \& {Diehl} et~al.,}{{Flaugher}
  et~al.}{2015}]{Flaugher2015}
{Flaugher} B.,  et~al. 2015, \mn@doi [\aj] {10.1088/0004-6256/150/5/150}, \href
  {https://ui.adsabs.harvard.edu/abs/2015AJ....150..150F} {150, 150}

\bibitem[\protect\citeauthoryear{{Foreman-Mackey} \& {Hogg}
  et~al.,}{{Foreman-Mackey} et~al.}{2013}]{Foreman-Mackey2013}
{Foreman-Mackey} D.,  et~al. 2013, \mn@doi [\pasp] {10.1086/670067}, \href
  {https://ui.adsabs.harvard.edu/abs/2013PASP..125..306F} {125, 306}

\bibitem[\protect\citeauthoryear{{Gatti} \& {Sheldon} et~al.,}{{Gatti}
  et~al.}{2021}]{Gatti2021}
{Gatti} M.,  et~al. 2021, \mn@doi [\mnras] {10.1093/mnras/stab918}, \href
  {https://ui.adsabs.harvard.edu/abs/2021MNRAS.504.4312G} {504, 4312}

\bibitem[\protect\citeauthoryear{{Giannini} \& {Alarcon} et~al.,}{{Giannini}
  et~al.}{2024}]{Giannini2022}
{Giannini} G.,  et~al. 2024, \mn@doi [\mnras] {10.1093/mnras/stad2945}, \href
  {https://ui.adsabs.harvard.edu/abs/2024MNRAS.527.2010G} {527, 2010}

\bibitem[\protect\citeauthoryear{{Girelli} \& {Pozzetti} et~al.,}{{Girelli}
  et~al.}{2020}]{Girelli2020}
{Girelli} G.,  et~al. 2020, \mn@doi [\aap] {10.1051/0004-6361/201936329}, \href
  {https://ui.adsabs.harvard.edu/abs/2020A&A...634A.135G} {634, A135}

\bibitem[\protect\citeauthoryear{{Goodman} \& {Weare}}{{Goodman} \&
  {Weare}}{2010}]{Goodman2010}
{Goodman} J.,  {Weare} J.,  2010, \mn@doi [Communications in Applied
  Mathematics and Computational Science] {10.2140/camcos.2010.5.65}, \href
  {https://ui.adsabs.harvard.edu/abs/2010CAMCS...5...65G} {5, 65}

\bibitem[\protect\citeauthoryear{{Greco} \& {Greene} et~al.,}{{Greco}
  et~al.}{2018}]{Greco2018}
{Greco} J.~P.,  et~al. 2018, \mn@doi [\apj] {10.3847/1538-4357/aab842}, \href
  {https://ui.adsabs.harvard.edu/abs/2018ApJ...857..104G} {857, 104}

\bibitem[\protect\citeauthoryear{{Greene} \& {Greco} et~al.,}{{Greene}
  et~al.}{2022}]{Greene2022}
{Greene} J.~E.,  et~al. 2022, \mn@doi [\apj] {10.3847/1538-4357/ac7238}, \href
  {https://ui.adsabs.harvard.edu/abs/2022ApJ...933..150G} {933, 150}

\bibitem[\protect\citeauthoryear{{Hambly} \& {MacGillivray} et~al.,}{{Hambly}
  et~al.}{2001}]{Hambly2001}
{Hambly} N.~C.,  et~al. 2001, \mn@doi [\mnras]
  {10.1111/j.1365-2966.2001.04660.x}, \href
  {https://ui.adsabs.harvard.edu/abs/2001MNRAS.326.1279H} {326, 1279}

\bibitem[\protect\citeauthoryear{{Hartlap}, {Simon}  \& {Schneider}}{{Hartlap}
  et~al.}{2007}]{Hartlap2007}
{Hartlap} J.,  {Simon} P.,   {Schneider} P.,  2007, \mn@doi [\aap]
  {10.1051/0004-6361:20066170}, \href
  {https://ui.adsabs.harvard.edu/abs/2007A&A...464..399H} {464, 399}

\bibitem[\protect\citeauthoryear{{Hayward}, {Irwin}  \& {Bregman}}{{Hayward}
  et~al.}{2005}]{Hayward2005}
{Hayward} C.~C.,  {Irwin} J.~A.,   {Bregman} J.~N.,  2005, \mn@doi [\apj]
  {10.1086/497565}, \href
  {https://ui.adsabs.harvard.edu/abs/2005ApJ...635..827H} {635, 827}

\bibitem[\protect\citeauthoryear{{Hinton}}{{Hinton}}{2016}]{Hinton2016}
{Hinton} S.~R.,  2016, \mn@doi [The Journal of Open Source Software]
  {10.21105/joss.00045}, \href
  {https://ui.adsabs.harvard.edu/abs/2016JOSS....1...45H} {1, 00045}

\bibitem[\protect\citeauthoryear{{Into} \& {Portinari}}{{Into} \&
  {Portinari}}{2013}]{Into2013}
{Into} T.,  {Portinari} L.,  2013, \mn@doi [\mnras] {10.1093/mnras/stt071},
  \href {https://ui.adsabs.harvard.edu/abs/2013MNRAS.430.2715I} {430, 2715}

\bibitem[\protect\citeauthoryear{{Ishiyama} \& {Prada} et~al.,}{{Ishiyama}
  et~al.}{2021}]{Ishiyama2021}
{Ishiyama} T.,  et~al. 2021, \mn@doi [\mnras] {10.1093/mnras/stab1755}, \href
  {https://ui.adsabs.harvard.edu/abs/2021MNRAS.506.4210I} {506, 4210}

\bibitem[\protect\citeauthoryear{{Ivezi{\'c}} \& {Kahn} et~al.,}{{Ivezi{\'c}}
  et~al.}{2019}]{Ivezic2019}
{Ivezi{\'c}} {\v{Z}}.,  et~al. 2019, \mn@doi [\apj] {10.3847/1538-4357/ab042c},
  \href {https://ui.adsabs.harvard.edu/abs/2019ApJ...873..111I} {873, 111}

\bibitem[\protect\citeauthoryear{James \& Roos}{James \&
  Roos}{1975}]{James1975}
James F.,  Roos M.,  1975, \mn@doi [Comput. Phys. Commun.]
  {10.1016/0010-4655(75)90039-9}, 10, 343

\bibitem[\protect\citeauthoryear{{Jarrett} \& {Chester} et~al.,}{{Jarrett}
  et~al.}{2000}]{Jarrett2000}
{Jarrett} T.~H.,  et~al. 2000, \mn@doi [\aj] {10.1086/301330}, \href
  {https://ui.adsabs.harvard.edu/abs/2000AJ....119.2498J} {119, 2498}

\bibitem[\protect\citeauthoryear{{Jarvis}, {Bernstein}  \& {Jain}}{{Jarvis}
  et~al.}{2004}]{Jarvis2004}
{Jarvis} M.,  {Bernstein} G.,   {Jain} B.,  2004, \mn@doi [\mnras]
  {10.1111/j.1365-2966.2004.07926.x}, \href
  {https://ui.adsabs.harvard.edu/abs/2004MNRAS.352..338J} {352, 338}

\bibitem[\protect\citeauthoryear{{Jiang} \& {Dekel} et~al.,}{{Jiang}
  et~al.}{2021}]{Jiang2021}
{Jiang} F.,  et~al. 2021, \mn@doi [\mnras] {10.1093/mnras/staa4034}, \href
  {https://ui.adsabs.harvard.edu/abs/2021MNRAS.502..621J} {502, 621}

\bibitem[\protect\citeauthoryear{{Kado-Fong} \& {Petrescu} et~al.,}{{Kado-Fong}
  et~al.}{2021}]{Kado-Fong2021}
{Kado-Fong} E.,  et~al. 2021, \mn@doi [\apj] {10.3847/1538-4357/ac15f0}, \href
  {https://ui.adsabs.harvard.edu/abs/2021ApJ...920...72K} {920, 72}

\bibitem[\protect\citeauthoryear{Kaufman}{Kaufman}{1967}]{Kaufman1967}
Kaufman G.~M.,  1967, Center for Operations Research and Econometrics
  Discussion Paper, 6710, 44

\bibitem[\protect\citeauthoryear{{Koda} \& {Yagi} et~al.,}{{Koda}
  et~al.}{2015}]{Koda2015}
{Koda} J.,  et~al. 2015, \mn@doi [\apjl] {10.1088/2041-8205/807/1/L2}, \href
  {https://ui.adsabs.harvard.edu/abs/2015ApJ...807L...2K} {807, L2}

\bibitem[\protect\citeauthoryear{{Kovacs} \& {Szapudi} et~al.,}{{Kovacs}
  et~al.}{2013}]{Kovacs2013}
{Kovacs} A.,  et~al. 2013, \mn@doi [\mnras] {10.1093/mnrasl/slt002}, \href
  {https://ui.adsabs.harvard.edu/abs/2013MNRAS.431L..28K} {431, L28}

\bibitem[\protect\citeauthoryear{{Kov{\'a}cs}, {Bogd{\'a}n}  \&
  {Canning}}{{Kov{\'a}cs} et~al.}{2019}]{Kovacs2019}
{Kov{\'a}cs} O.~E.,  {Bogd{\'a}n} {\'A}.,   {Canning} R. E.~A.,  2019, \mn@doi
  [\apjl] {10.3847/2041-8213/ab2916}, \href
  {https://ui.adsabs.harvard.edu/abs/2019ApJ...879L..12K} {879, L12}

\bibitem[\protect\citeauthoryear{{Leauthaud} \& {Singh} et~al.,}{{Leauthaud}
  et~al.}{2020}]{Leauthaud2020}
{Leauthaud} A.,  et~al. 2020, \mn@doi [Physics of the Dark Universe]
  {10.1016/j.dark.2020.100719}, \href
  {https://ui.adsabs.harvard.edu/abs/2020PDU....3000719L} {30, 100719}

\bibitem[\protect\citeauthoryear{{Lee} \& {Kang} et~al.,}{{Lee}
  et~al.}{2018}]{Lee2018}
{Lee} M.~G.,  et~al. 2018, \mn@doi [VizieR Online Data Catalog]
  {10.26093/cds/vizier.18440157}, \href
  {https://ui.adsabs.harvard.edu/abs/2018yCat..18440157L} {p. J/ApJ/844/157}

\bibitem[\protect\citeauthoryear{{Leisman} \& {Haynes} et~al.,}{{Leisman}
  et~al.}{2017}]{Leisman2017}
{Leisman} L.,  et~al. 2017, \mn@doi [\apj] {10.3847/1538-4357/aa7575}, \href
  {https://ui.adsabs.harvard.edu/abs/2017ApJ...842..133L} {842, 133}

\bibitem[\protect\citeauthoryear{{Li} \& {Mo} et~al.,}{{Li}
  et~al.}{2013}]{Li2013}
{Li} R.,  et~al. 2013, \mn@doi [\mnras] {10.1093/mnras/stt133}, \href
  {https://ui.adsabs.harvard.edu/abs/2013MNRAS.430.3359L} {430, 3359}

\bibitem[\protect\citeauthoryear{{Ludlow} \& {Bose} et~al.,}{{Ludlow}
  et~al.}{2016}]{Ludlow2016}
{Ludlow} A.~D.,  et~al. 2016, \mn@doi [\mnras] {10.1093/mnras/stw1046}, \href
  {https://ui.adsabs.harvard.edu/abs/2016MNRAS.460.1214L} {460, 1214}

\bibitem[\protect\citeauthoryear{{Luo} \& {Leauthaud} et~al.,}{{Luo}
  et~al.}{2024}]{Luo2023}
{Luo} Y.,  et~al. 2024, \mn@doi [\mnras] {10.1093/mnras/stae925}, \href
  {https://ui.adsabs.harvard.edu/abs/2024MNRAS.530.4988L} {530, 4988}

\bibitem[\protect\citeauthoryear{{Madhavacheril} \& {Sif{\'o}n}
  et~al.,}{{Madhavacheril} et~al.}{2020}]{Madhavacheril2020}
{Madhavacheril} M.~S.,  et~al. 2020, \mn@doi [\apjl]
  {10.3847/2041-8213/abbccb}, \href
  {https://ui.adsabs.harvard.edu/abs/2020ApJ...903L..13M} {903, L13}

\bibitem[\protect\citeauthoryear{{Mandelbaum}}{{Mandelbaum}}{2015}]{Mandelbaum2015}
{Mandelbaum} R.,  2015, in {Cappellari} M.,  {Courteau} S.,  eds, ~ Vol. 311,
  Galaxy Masses as Constraints of Formation Models. pp 86--95 (\mn@eprint
  {arXiv} {1410.0734}), \mn@doi{10.1017/S1743921315003452}

\bibitem[\protect\citeauthoryear{{Mao} \& {Geha} et~al.,}{{Mao}
  et~al.}{2021}]{Mao2021}
{Mao} Y.-Y.,  et~al. 2021, \mn@doi [\apj] {10.3847/1538-4357/abce58}, \href
  {https://ui.adsabs.harvard.edu/abs/2021ApJ...907...85M} {907, 85}

\bibitem[\protect\citeauthoryear{{Martin} \& {Ibata} et~al.,}{{Martin}
  et~al.}{2013}]{Martin2013}
{Martin} N.~F.,  et~al. 2013, \mn@doi [\apj] {10.1088/0004-637X/776/2/80},
  \href {https://ui.adsabs.harvard.edu/abs/2013ApJ...776...80M} {776, 80}

\bibitem[\protect\citeauthoryear{{Martin} \& {Kaviraj} et~al.,}{{Martin}
  et~al.}{2019}]{Martin2019}
{Martin} G.,  et~al. 2019, \mn@doi [\mnras] {10.1093/mnras/stz356}, \href
  {https://ui.adsabs.harvard.edu/abs/2019MNRAS.485..796M} {485, 796}

\bibitem[\protect\citeauthoryear{{Mart{\'\i}nez-Delgado} \& {L{\"a}sker}
  et~al.,}{{Mart{\'\i}nez-Delgado} et~al.}{2016}]{Martinez-Delgado2016}
{Mart{\'\i}nez-Delgado} D.,  et~al. 2016, \mn@doi [\aj]
  {10.3847/0004-6256/151/4/96}, \href
  {https://ui.adsabs.harvard.edu/abs/2016AJ....151...96M} {151, 96}

\bibitem[\protect\citeauthoryear{{McConnachie}}{{McConnachie}}{2012}]{McConnachie2012}
{McConnachie} A.~W.,  2012, \mn@doi [\aj] {10.1088/0004-6256/144/1/4}, \href
  {https://ui.adsabs.harvard.edu/abs/2012AJ....144....4M} {144, 4}

\bibitem[\protect\citeauthoryear{{McEnery}}{{McEnery}}{2021}]{McEnery2021}
{McEnery} J.,  2021, in American Astronomical Society Meeting Abstracts. p.
  327.01

\bibitem[\protect\citeauthoryear{{McGaugh}}{{McGaugh}}{1996}]{McGaugh1996}
{McGaugh} S.~S.,  1996, \mn@doi [\mnras] {10.1093/mnras/280.2.337}, \href
  {https://ui.adsabs.harvard.edu/abs/1996MNRAS.280..337M} {280, 337}

\bibitem[\protect\citeauthoryear{{M{\'e}nard} \& {Scranton}
  et~al.,}{{M{\'e}nard} et~al.}{2013}]{Menard2013}
{M{\'e}nard} B.,  et~al. 2013, \mn@doi [arXiv e-prints]
  {10.48550/arXiv.1303.4722}, \href
  {https://ui.adsabs.harvard.edu/abs/2013arXiv1303.4722M} {p. arXiv:1303.4722}

\bibitem[\protect\citeauthoryear{{Mihos} \& {Durrell} et~al.,}{{Mihos}
  et~al.}{2015}]{Mihos2015}
{Mihos} J.~C.,  et~al. 2015, \mn@doi [\apjl] {10.1088/2041-8205/809/2/L21},
  \href {https://ui.adsabs.harvard.edu/abs/2015ApJ...809L..21M} {809, L21}

\bibitem[\protect\citeauthoryear{{Minchin} \& {Disney} et~al.,}{{Minchin}
  et~al.}{2004}]{Minchin2004}
{Minchin} R.~F.,  et~al. 2004, \mn@doi [\mnras]
  {10.1111/j.1365-2966.2004.08409.x}, \href
  {https://ui.adsabs.harvard.edu/abs/2004MNRAS.355.1303M} {355, 1303}

\bibitem[\protect\citeauthoryear{{Moster} \& {Somerville} et~al.,}{{Moster}
  et~al.}{2010}]{Moster2010}
{Moster} B.~P.,  et~al. 2010, \mn@doi [\apj] {10.1088/0004-637X/710/2/903},
  \href {https://ui.adsabs.harvard.edu/abs/2010ApJ...710..903M} {710, 903}

\bibitem[\protect\citeauthoryear{{Mu{\~n}oz} \& {Eigenthaler}
  et~al.,}{{Mu{\~n}oz} et~al.}{2015}]{Munoz2015}
{Mu{\~n}oz} R.~P.,  et~al. 2015, \mn@doi [\apjl] {10.1088/2041-8205/813/1/L15},
  \href {https://ui.adsabs.harvard.edu/abs/2015ApJ...813L..15M} {813, L15}

\bibitem[\protect\citeauthoryear{{Munshi} \& {Brooks} et~al.,}{{Munshi}
  et~al.}{2021}]{Munshi2021}
{Munshi} F.,  et~al. 2021, \mn@doi [\apj] {10.3847/1538-4357/ac0db6}, \href
  {https://ui.adsabs.harvard.edu/abs/2021ApJ...923...35M} {923, 35}

\bibitem[\protect\citeauthoryear{{Myles} \& {Alarcon} et~al.,}{{Myles}
  et~al.}{2021}]{Myles2021}
{Myles} J.,  et~al. 2021, \mn@doi [\mnras] {10.1093/mnras/stab1515}, \href
  {https://ui.adsabs.harvard.edu/abs/2021MNRAS.505.4249M} {505, 4249}

\bibitem[\protect\citeauthoryear{{Navarro}, {Frenk}  \& {White}}{{Navarro}
  et~al.}{1996}]{Navarro1996}
{Navarro} J.~F.,  {Frenk} C.~S.,   {White} S. D.~M.,  1996, \mn@doi [\apj]
  {10.1086/177173}, \href
  {https://ui.adsabs.harvard.edu/abs/1996ApJ...462..563N} {462, 563}

\bibitem[\protect\citeauthoryear{{Niemiec} \& {Giocoli} et~al.,}{{Niemiec}
  et~al.}{2022}]{Niemiec2022}
{Niemiec} A.,  et~al. 2022, \mn@doi [\mnras] {10.1093/mnras/stac832}, \href
  {https://ui.adsabs.harvard.edu/abs/2022MNRAS.512.6021N} {512, 6021}

\bibitem[\protect\citeauthoryear{{O'Neil} \& {Bothun}}{{O'Neil} \&
  {Bothun}}{2000}]{ONeil2000}
{O'Neil} K.,  {Bothun} G.,  2000, \mn@doi [\apj] {10.1086/308322}, \href
  {https://ui.adsabs.harvard.edu/abs/2000ApJ...529..811O} {529, 811}

\bibitem[\protect\citeauthoryear{{Pandey} \& {Krause} et~al.,}{{Pandey}
  et~al.}{2022}]{Pandey2022}
{Pandey} S.,  et~al. 2022, \mn@doi [\prd] {10.1103/PhysRevD.106.043520}, \href
  {https://ui.adsabs.harvard.edu/abs/2022PhRvD.106d3520P} {106, 043520}

\bibitem[\protect\citeauthoryear{{Planck Collaboration} \& {Ade}
  et~al.,}{{Planck Collaboration}}{2016}]{Planck2016}
{Planck Collaboration} 2016, \mn@doi [\aap] {10.1051/0004-6361/201525830},
  \href {https://ui.adsabs.harvard.edu/abs/2016A&A...594A..13P} {594, A13}

\bibitem[\protect\citeauthoryear{{Porredon} \& {Crocce} et~al.,}{{Porredon}
  et~al.}{2022}]{Porredon2022}
{Porredon} A.,  et~al. 2022, \mn@doi [\prd] {10.1103/PhysRevD.106.103530},
  \href {https://ui.adsabs.harvard.edu/abs/2022PhRvD.106j3530P} {106, 103530}

\bibitem[\protect\citeauthoryear{{Prat} \& {S{\'a}nchez} et~al.,}{{Prat}
  et~al.}{2018}]{Prat2018}
{Prat} J.,  et~al. 2018, \mn@doi [\prd] {10.1103/PhysRevD.98.042005}, \href
  {https://ui.adsabs.harvard.edu/abs/2018PhRvD..98d2005P} {98, 042005}

\bibitem[\protect\citeauthoryear{{Prat} \& {Blazek} et~al.,}{{Prat}
  et~al.}{2022}]{Prat2022}
{Prat} J.,  et~al. 2022, \mn@doi [\prd] {10.1103/PhysRevD.105.083528}, \href
  {https://ui.adsabs.harvard.edu/abs/2022PhRvD.105h3528P} {105, 083528}

\bibitem[\protect\citeauthoryear{{Prole} \& {Hilker} et~al.,}{{Prole}
  et~al.}{2019}]{Prole2019}
{Prole} D.~J.,  et~al. 2019, \mn@doi [\mnras] {10.1093/mnras/stz326}, \href
  {https://ui.adsabs.harvard.edu/abs/2019MNRAS.484.4865P} {484, 4865}

\bibitem[\protect\citeauthoryear{{Prole} \& {van der Burg} et~al.,}{{Prole}
  et~al.}{2021}]{Prole2021}
{Prole} D.~J.,  et~al. 2021, \mn@doi [\mnras] {10.1093/mnras/staa3296}, \href
  {https://ui.adsabs.harvard.edu/abs/2021MNRAS.500.2049P} {500, 2049}

\bibitem[\protect\citeauthoryear{{Rozo} \& {Rykoff} et~al.,}{{Rozo}
  et~al.}{2016}]{Rozo2016}
{Rozo} E.,  et~al. 2016, \mn@doi [\mnras] {10.1093/mnras/stw1281}, \href
  {https://ui.adsabs.harvard.edu/abs/2016MNRAS.461.1431R} {461, 1431}

\bibitem[\protect\citeauthoryear{{Schmidt} \& {M{\'e}nard} et~al.,}{{Schmidt}
  et~al.}{2013}]{Schmidt2013}
{Schmidt} S.~J.,  et~al. 2013, \mn@doi [\mnras] {10.1093/mnras/stt410}, \href
  {https://ui.adsabs.harvard.edu/abs/2013MNRAS.431.3307S} {431, 3307}

\bibitem[\protect\citeauthoryear{{Sevilla-Noarbe} \& {Bechtol}
  et~al.,}{{Sevilla-Noarbe} et~al.}{2021}]{Sevilla-Noarbe2021}
{Sevilla-Noarbe} I.,  et~al. 2021, \mn@doi [\apjs] {10.3847/1538-4365/abeb66},
  \href {https://ui.adsabs.harvard.edu/abs/2021ApJS..254...24S} {254, 24}

\bibitem[\protect\citeauthoryear{{Sheldon} \& {Huff}}{{Sheldon} \&
  {Huff}}{2017}]{Sheldon2017}
{Sheldon} E.~S.,  {Huff} E.~M.,  2017, \mn@doi [\apj]
  {10.3847/1538-4357/aa704b}, \href
  {https://ui.adsabs.harvard.edu/abs/2017ApJ...841...24S} {841, 24}

\bibitem[\protect\citeauthoryear{{Sheldon} \& {Johnston} et~al.,}{{Sheldon}
  et~al.}{2004}]{Sheldon2004}
{Sheldon} E.~S.,  et~al. 2004, \mn@doi [\aj] {10.1086/383293}, \href
  {https://ui.adsabs.harvard.edu/abs/2004AJ....127.2544S} {127, 2544}

\bibitem[\protect\citeauthoryear{{Sif{\'o}n} \& {van der Burg}
  et~al.,}{{Sif{\'o}n} et~al.}{2018a}]{Sifon2021}
{Sif{\'o}n} C.,  et~al. 2018a, \mn@doi [\mnras] {10.1093/mnras/stx2648}, \href
  {https://ui.adsabs.harvard.edu/abs/2018MNRAS.473.3747S} {473, 3747}

\bibitem[\protect\citeauthoryear{{Sif{\'o}n} \& {Herbonnet}
  et~al.,}{{Sif{\'o}n} et~al.}{2018b}]{Sifon2018}
{Sif{\'o}n} C.,  et~al. 2018b, \mn@doi [\mnras] {10.1093/mnras/sty1161}, \href
  {https://ui.adsabs.harvard.edu/abs/2018MNRAS.478.1244S} {478, 1244}

\bibitem[\protect\citeauthoryear{{Singh} \& {Mandelbaum} et~al.,}{{Singh}
  et~al.}{2017}]{Singh2017}
{Singh} S.,  et~al. 2017, \mn@doi [\mnras] {10.1093/mnras/stx1828}, \href
  {https://ui.adsabs.harvard.edu/abs/2017MNRAS.471.3827S} {471, 3827}

\bibitem[\protect\citeauthoryear{{Takahashi} \& {Sato} et~al.,}{{Takahashi}
  et~al.}{2012}]{Takahashi2012}
{Takahashi} R.,  et~al. 2012, \mn@doi [\apj] {10.1088/0004-637X/761/2/152},
  \href {https://ui.adsabs.harvard.edu/abs/2012ApJ...761..152T} {761, 152}

\bibitem[\protect\citeauthoryear{{Tanoglidis} \& {Drlica-Wagner}
  et~al.,}{{Tanoglidis} et~al.}{2021}]{Tanoglidis2021}
{Tanoglidis} D.,  et~al. 2021, \mn@doi [\apjs] {10.3847/1538-4365/abca89},
  \href {https://ui.adsabs.harvard.edu/abs/2021ApJS..252...18T} {252, 18}

\bibitem[\protect\citeauthoryear{{Thornton} \& {Amon} et~al.,}{{Thornton}
  et~al.}{2023}]{Thornton2023}
{Thornton} J.,  et~al. 2023, \mn@doi [arXiv e-prints]
  {10.48550/arXiv.2311.14659}, \href
  {https://ui.adsabs.harvard.edu/abs/2023arXiv231114659T} {p. arXiv:2311.14659}

\bibitem[\protect\citeauthoryear{{Thuruthipilly} \& {Junais}
  et~al.,}{{Thuruthipilly} et~al.}{2024}]{Thuruthipilly2023}
{Thuruthipilly} H.,  et~al. 2024, \mn@doi [\aap] {10.1051/0004-6361/202347649},
  \href {https://ui.adsabs.harvard.edu/abs/2024A&A...682A...4T} {682, A4}

\bibitem[\protect\citeauthoryear{{Wang} \& {Li} et~al.,}{{Wang}
  et~al.}{2024}]{wang2024}
{Wang} C.,  et~al. 2024, \mn@doi [\mnras] {10.1093/mnras/stae121}, \href
  {https://ui.adsabs.harvard.edu/abs/2024MNRAS.528.2728W} {528, 2728}

\bibitem[\protect\citeauthoryear{{Yagi} \& {Koda} et~al.,}{{Yagi}
  et~al.}{2016}]{Yagi2016}
{Yagi} M.,  et~al. 2016, \mn@doi [\apjs] {10.3847/0067-0049/225/1/11}, \href
  {https://ui.adsabs.harvard.edu/abs/2016ApJS..225...11Y} {225, 11}

\bibitem[\protect\citeauthoryear{{Zacharegkas} \& {Chang}
  et~al.,}{{Zacharegkas} et~al.}{2022}]{Zacharegkas2022}
{Zacharegkas} G.,  et~al. 2022, \mn@doi [\mnras] {10.1093/mnras/stab3155},
  \href {https://ui.adsabs.harvard.edu/abs/2022MNRAS.509.3119Z} {509, 3119}

\bibitem[\protect\citeauthoryear{{Zehavi} \& {Zheng} et~al.,}{{Zehavi}
  et~al.}{2011}]{Zehavi2011}
{Zehavi} I.,  et~al. 2011, \mn@doi [\apj] {10.1088/0004-637X/736/1/59}, \href
  {https://ui.adsabs.harvard.edu/abs/2011ApJ...736...59Z} {736, 59}

\bibitem[\protect\citeauthoryear{{Zuntz} \& {Paterno} et~al.,}{{Zuntz}
  et~al.}{2015}]{Zuntz2015}
{Zuntz} J.,  et~al. 2015, \mn@doi [Astronomy and Computing]
  {10.1016/j.ascom.2015.05.005}, \href
  {https://ui.adsabs.harvard.edu/abs/2015A&C....12...45Z} {12, 45}

\bibitem[\protect\citeauthoryear{{van Dokkum} \& {Abraham} et~al.,}{{van
  Dokkum} et~al.}{2016}]{vanDokkum2016}
{van Dokkum} P.,  et~al. 2016, \mn@doi [\apjl] {10.3847/2041-8205/828/1/L6},
  \href {https://ui.adsabs.harvard.edu/abs/2016ApJ...828L...6V} {828, L6}

\bibitem[\protect\citeauthoryear{{van der Burg}, {Muzzin}  \& {Hoekstra}}{{van
  der Burg} et~al.}{2016}]{vanderBurgh2016}
{van der Burg} R. F.~J.,  {Muzzin} A.,   {Hoekstra} H.,  2016, \mn@doi [\aap]
  {10.1051/0004-6361/201628222}, \href
  {https://ui.adsabs.harvard.edu/abs/2016A&A...590A..20V} {590, A20}

\makeatother
\end{thebibliography}
\bibliographystyle{mnras_2author}

\section*{Affiliations}


$^{1}$ Department of Astronomy and Astrophysics, University of Chicago, Chicago, IL 60637, USA\\
$^{2}$ Nordita, KTH Royal Institute of Technology and Stockholm University, Hannes Alfv\'ens v\"ag 12, SE-10691 Stockholm, Sweden\\
$^{3}$ Kavli Institute for Cosmological Physics, University of Chicago, Chicago, IL 60637, USA\\
$^{4}$ Fermi National Accelerator Laboratory, P. O. Box 500, Batavia, IL 60510, USA\\
$^{5}$ Indian Institute of Science Education and Research, Pune, Dr. Homi Babha Road, Pashan, 411008\\
$^{6}$ Department of Astrophysical Sciences, Princeton University, Peyton Hall, Princeton, NJ 08544, USA\\
$^{7}$ Department of Physics, Stanford University, 382 Via Pueblo Mall, Stanford, CA 94305, USA\\
$^{8}$ Kavli Institute for Particle Astrophysics \& Cosmology, P. O. Box 2450, Stanford University, Stanford, CA 94305, USA\\
$^{9}$ SLAC National Accelerator Laboratory, Menlo Park, CA 94025, USA\\
$^{10}$ Argonne National Laboratory, 9700 South Cass Avenue, Lemont, IL 60439, USA\\
$^{11}$ Institute of Space Sciences (ICE, CSIC),  Campus UAB, Carrer de Can Magrans, s/n,  08193 Barcelona, Spain\\
$^{12}$ Physics Department, 2320 Chamberlin Hall, University of Wisconsin-Madison, 1150 University Avenue Madison, WI  53706-1390\\
$^{13}$ Department of Physics and Astronomy, University of Pennsylvania, Philadelphia, PA 19104, USA\\
$^{14}$ Department of Physics, Carnegie Mellon University, Pittsburgh, Pennsylvania 15312, USA\\
$^{15}$ NSF AI Planning Institute for Physics of the Future, Carnegie Mellon University, Pittsburgh, PA 15213, USA\\
$^{16}$ Instituto de Astrofisica de Canarias, E-38205 La Laguna, Tenerife, Spain\\
$^{17}$ Laborat\'orio Interinstitucional de e-Astronomia - LIneA, Rua Gal. Jos\'e Cristino 77, Rio de Janeiro, RJ - 20921-400, Brazil\\
$^{18}$ Center for Astrophysical Surveys, National Center for Supercomputing Applications, 1205 West Clark St., Urbana, IL 61801, USA\\
$^{19}$ Department of Astronomy, University of Illinois at Urbana-Champaign, 1002 W. Green Street, Urbana, IL 61801, USA\\
$^{20}$ Physics Department, William Jewell College, Liberty, MO, 64068\\
$^{21}$ Department of Physics, Duke University Durham, NC 27708, USA\\
$^{22}$ NASA Goddard Space Flight Center, 8800 Greenbelt Rd, Greenbelt, MD 20771, USA\\
$^{23}$ Jodrell Bank Center for Astrophysics, School of Physics and Astronomy, University of Manchester, Oxford Road, Manchester, M13 9PL, UK\\
$^{24}$ Lawrence Berkeley National Laboratory, 1 Cyclotron Road, Berkeley, CA 94720, USA\\
$^{25}$ Universit\'e Grenoble Alpes, CNRS, LPSC-IN2P3, 38000 Grenoble, France\\
$^{26}$ Department of Physics and Astronomy, University of Waterloo, 200 University Ave W, Waterloo, ON N2L 3G1, Canada\\
$^{27}$ Jet Propulsion Laboratory, California Institute of Technology, 4800 Oak Grove Dr., Pasadena, CA 91109, USA\\
$^{28}$ Institut de F\'{\i}sica d'Altes Energies (IFAE), The Barcelona Institute of Science and Technology, Campus UAB, 08193 Bellaterra (Barcelona) Spain\\
$^{29}$ University Observatory, Faculty of Physics, Ludwig-Maximilians-Universit\"at, Scheinerstr. 1, 81679 Munich, Germany\\
$^{30}$ School of Physics and Astronomy, Cardiff University, CF24 3AA, UK\\
$^{31}$ Department of Applied Mathematics and Theoretical Physics, University of Cambridge, Cambridge CB3 0WA, UK\\
$^{32}$ Instituto de F\'isica Gleb Wataghin, Universidade Estadual de Campinas, 13083-859, Campinas, SP, Brazil\\
$^{33}$ Department of Physics, University of Genova and INFN, Via Dodecaneso 33, 16146, Genova, Italy\\
$^{34}$ Center for Cosmology and Astro-Particle Physics, The Ohio State University, Columbus, OH 43210, USA\\
$^{35}$ Centro de Investigaciones Energ\'eticas, Medioambientales y Tecnol\'ogicas (CIEMAT), Madrid, Spain\\
$^{36}$ Brookhaven National Laboratory, Bldg 510, Upton, NY 11973, USA\\
$^{37}$ Department of Physics and Astronomy, Stony Brook University, Stony Brook, NY 11794, USA\\
$^{38}$ Institut de Recherche en Astrophysique et Plan\'etologie (IRAP), Universit\'e de Toulouse, CNRS, UPS, CNES, 14 Av. Edouard Belin, 31400 Toulouse, France\\
$^{39}$ Excellence Cluster Origins, Boltzmannstr.\ 2, 85748 Garching, Germany\\
$^{40}$ Max Planck Institute for Extraterrestrial Physics, Giessenbachstrasse, 85748 Garching, Germany\\
$^{41}$ Universit\"ats-Sternwarte, Fakult\"at f\"ur Physik, Ludwig-Maximilians Universit\"at M\"unchen, Scheinerstr. 1, 81679 M\"unchen, Germany\\
$^{42}$ Institute for Astronomy, University of Edinburgh, Edinburgh EH9 3HJ, UK\\
$^{43}$ Department of Physics, University of Michigan, Ann Arbor, MI 48109, USA\\
$^{44}$ Institute of Cosmology and Gravitation, University of Portsmouth, Portsmouth, PO1 3FX, UK\\
$^{45}$ Department of Physics \& Astronomy, University College London, Gower Street, London, WC1E 6BT, UK\\
$^{46}$ Institut d'Estudis Espacials de Catalunya (IEEC), 08034 Barcelona, Spain\\
$^{47}$ University of Nottingham, School of Physics and Astronomy, Nottingham NG7 2RD, UK\\
$^{48}$ Department of Physics, IIT Hyderabad, Kandi, Telangana 502285, India\\
$^{49}$ Institute of Theoretical Astrophysics, University of Oslo. P.O. Box 1029 Blindern, NO-0315 Oslo, Norway\\
$^{50}$ Instituto de Fisica Teorica UAM/CSIC, Universidad Autonoma de Madrid, 28049 Madrid, Spain\\
$^{51}$ School of Mathematics and Physics, University of Queensland,  Brisbane, QLD 4072, Australia\\
$^{52}$ Santa Cruz Institute for Particle Physics, Santa Cruz, CA 95064, USA\\
$^{53}$ Department of Physics, The Ohio State University, Columbus, OH 43210, USA\\
$^{54}$ Center for Astrophysics $\vert$ Harvard \& Smithsonian, 60 Garden Street, Cambridge, MA 02138, USA\\
$^{55}$ Australian Astronomical Optics, Macquarie University, North Ryde, NSW 2113, Australia\\
$^{56}$ Lowell Observatory, 1400 Mars Hill Rd, Flagstaff, AZ 86001, USA\\
$^{57}$ Centre for Gravitational Astrophysics, College of Science, The Australian National University, ACT 2601, Australia\\
$^{58}$ The Research School of Astronomy and Astrophysics, Australian National University, ACT 2601, Australia\\
$^{59}$ Departamento de F\'isica Matem\'atica, Instituto de F\'isica, Universidade de S\~ao Paulo, CP 66318, S\~ao Paulo, SP, 05314-970, Brazil\\
$^{60}$ George P. and Cynthia Woods Mitchell Institute for Fundamental Physics and Astronomy, and Department of Physics and Astronomy, Texas A\&M University, College Station, TX 77843,  USA\\
$^{61}$ LPSC Grenoble - 53, Avenue des Martyrs 38026 Grenoble, France\\
$^{62}$ Instituci\'o Catalana de Recerca i Estudis Avan\c{c}ats, E-08010 Barcelona, Spain\\
$^{63}$ Perimeter Institute for Theoretical Physics, 31 Caroline St. North, Waterloo, ON N2L 2Y5, Canada\\
$^{64}$ Observat\'orio Nacional, Rua Gal. Jos\'e Cristino 77, Rio de Janeiro, RJ - 20921-400, Brazil\\
$^{65}$ Hamburger Sternwarte, Universit\"{a}t Hamburg, Gojenbergsweg 112, 21029 Hamburg, Germany\\
$^{66}$ Ruhr University Bochum, Faculty of Physics and Astronomy, Astronomical Institute, German Centre for Cosmological Lensing, 44780 Bochum, Germany\\
$^{67}$ Cerro Tololo Inter-American Observatory, NSF's National Optical-Infrared Astronomy Research Laboratory, Casilla 603, La Serena, Chile\\
$^{68}$ Department of Physics, Northeastern University, Boston, MA 02115, USA\\
$^{69}$ School of Physics and Astronomy, University of Southampton,  Southampton, SO17 1BJ, UK\\
$^{70}$ Computer Science and Mathematics Division, Oak Ridge National Laboratory, Oak Ridge, TN 37831\\
$^{71}$ Department of Astronomy, University of California, Berkeley,  501 Campbell Hall, Berkeley, CA 94720, USA\\

\appendix \label{sec:appendix}
\section{Validation of the Data Vector}\label{sec:measurementvalidation}
\subsection{$\gamma_{\times}$ Shear}
\begin{figure}
    \centering
    \includegraphics[width=0.45\textwidth]{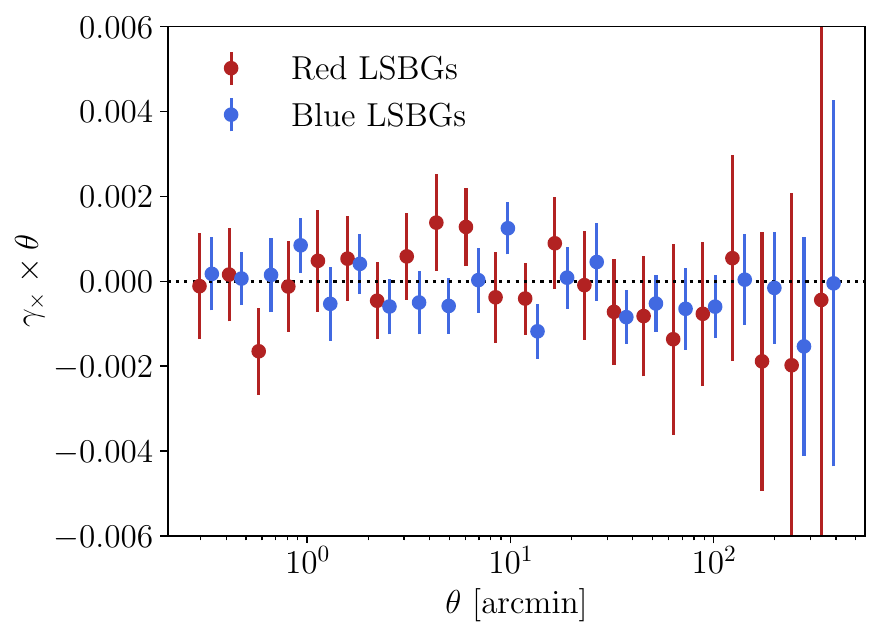}
    \caption{Measurement of the $\gamma _{\times} $ term, or the measurement of the lensing B-mode, for the red and blue LSBG samples. Note that the blue sample positions are offset from the red sample positions to improve visibility. We find this term to be consistent with zero.}
    \label{fig:gamma_cross}
\end{figure}

We compute the $\gamma_{\times} $ term as a systematics test on the tangential shear measurements. While the $\gamma_t$ term measures the shear of the lensing E-mode, the $ \gamma_{\times} $ term measures the shear of the lensing B-mode. We define $\gamma_{\times}$ as, 
\begin{equation}\label{eq:gamma_cross} 
e_{\times} = e_1\sin(2\phi) -e_2\cos(2\phi),
\end{equation} 
\noindent where $e_{\times}$ represents the cross component of the ellipticity and the other components follow the same definitions as Eq.~\ref{eq:ellipciticy}. Weak lensing only produces a tangential shear, thus in the absence of systematics the $\gamma_{\times}$ term must remain consistent with zero, as shown in Fig.~\ref{fig:gamma_cross}. Compared to a null model, we measure a $\chi ^2_\mathrm{null}$ for the $\gamma_{\times}$ term of 13.58/22 degrees of freedom for the full sample, 14.84/22 for the red sample, and 15.96/22 for the blue sample. These $\chi^2/\nu$ values are low, indicating overestimated uncertainties or biases drawn from the noisy inverted covariance.

\subsection{Random Sample Shear}
\begin{figure}
    \centering
    \includegraphics[width=0.45\textwidth]{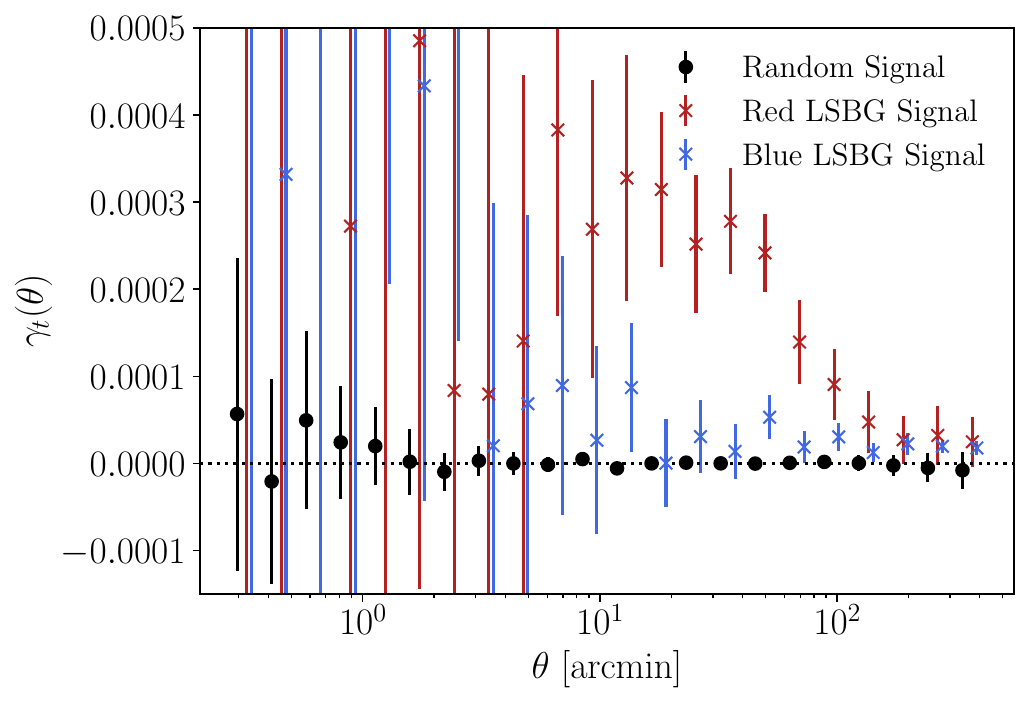}
    \caption{Tangential shear measurements around the random point sample, divided into 22 angular bins, compared to the red and blue LSBG lensing signals. All shape catalog redshift bins are combined. The shear measurements do not produce a significant signal, with a $\chi ^2_\mathrm{null}$ of 3.21/22.}
    \label{fig:random_shear_measurement}
\end{figure}

We measure the tangential shear surrounding the positions in the randoms catalog {and compare to the red and blue LSBG tangential shear signal.} These positions are randomly selected, therefore we should not measure any lensing signal. We find a $\chi^2_\mathrm{null}$ of 3.21 over 22 degrees of freedom, demonstrating a null signal. We note that for the blue LSBGs we remove scales of $\theta>50~\mathrm{arcmin}$ from our models. We illustrate this result in Fig.~\ref{fig:random_shear_measurement}.

\subsection{Boost Factor}
\begin{figure}
    \centering
    \includegraphics[width=0.45\textwidth]{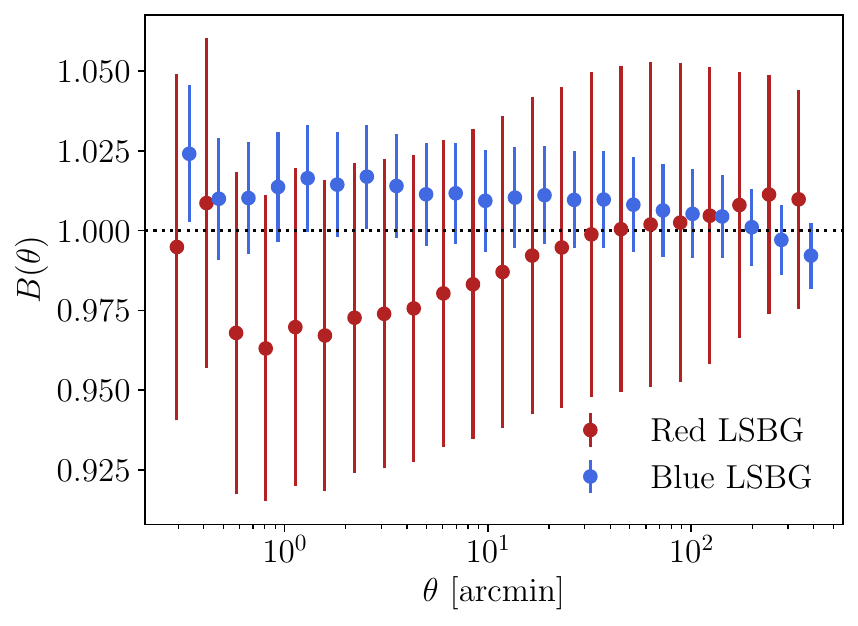}
    \caption{Boost factor measurements for the red and blue LSBG samples, with an offset in $\theta$ to improve visibility. The errorbars come from the jackknife covariance. The boost factor measurements are highly correlated and consistent with unity for both red and blue galaxies. The increased spatial variation between the red galaxy positions in the DES footprint heightens the boost factor covariance measurements, leading to larger errorbars.}
    \label{fig:boost_factor}
\end{figure}

The tangential shear measurements depend on the number density of the lens and source galaxies as a function of redshift, but ignore the clustering between lens and source galaxies. At small scales, this angular lens-source clustering can change the number of lens-source pairs when compared to predictions based on the mean number densities. As a result, the shear estimator becomes biased. The degree of bias depends on the scale. In other words, if a lens and a source galaxy live at the same redshift, lensing cannot occur. To estimate the decrease in lensing signal caused by lens-source clustering, we can compare the excess number of sources around lenses to the number of sources around random points as a function of scale. The resulting measurement is called the boost factor \citep{Sheldon2004}. We define the boost factor as \begin{equation} \label{eq:boostfactor} B(\theta) = 1 + \omega_{LS}(\theta) = \frac{\Sigma_R w_R}{\Sigma_L w_L}\frac{\Sigma_{LS}w_{LS}(\theta)}{\Sigma_{RS}w_{RS}(\theta)}.\end{equation} We include the ratio between the sum of random point weights and lens galaxy weights to normalize the boost factor. We plot the boost factor across angular scales in Fig.~\ref{fig:boost_factor} and find that the measurements are highly correlated and consistent with unity. This result implies that there isn't a significant redshift overlap between the lenses and sources, confirming a low-redshift lens system. To further ensure that any overlap is negligible, we test the impact of removing the first tomographic bin from the combined source redshift distribution and performing the tangential shear measurements. We find a $\Delta \chi^2$ of only 0.18 between the fiducial shear measurements and the measurements performed without the first tomographic redshift bin, indicating that our results are stable. The increased variation between the red galaxy spatial positions across the DES footprint likely heightens the size of the errorbars for the boost factor measurements by augmenting the differences between the jackknife patches used to calculate the covariance for the red LSBGs.

Including the boost factor in the tangential shear measurements results in a negligible $\Delta \chi^2$ of 0.009 for the red sample and 0.002 for the blue sample when compared to the fiducial tangential shear measurements. Moreover, we find the boost factor for the red sample to be smaller than one (while still compatible with unity), which indicates that other systematics, such as magnification, must dominate over the clustering between lenses and sources. Given the boost factor's sensitivity to these unmodeled systematics and its insignificant impact, we opt not to include it in the fiducial tangential shear measurements.

\section{Testing Robustness Against Priors}\label{sec:testingpriors}
In this appendix, we describe additional tests performed to check the robustness of our results against the chosen priors in Table~\ref{tab:model_params}. In particular, we test two scenarios (summarized in Table~\ref{tab:model_params_alt}): 
(1) marginalizing over the source redshift uncertainty, and (2) including the width of the offset distribution as a free parameter. 

\begin{figure}
    \centering
    \includegraphics[width=0.48\textwidth]{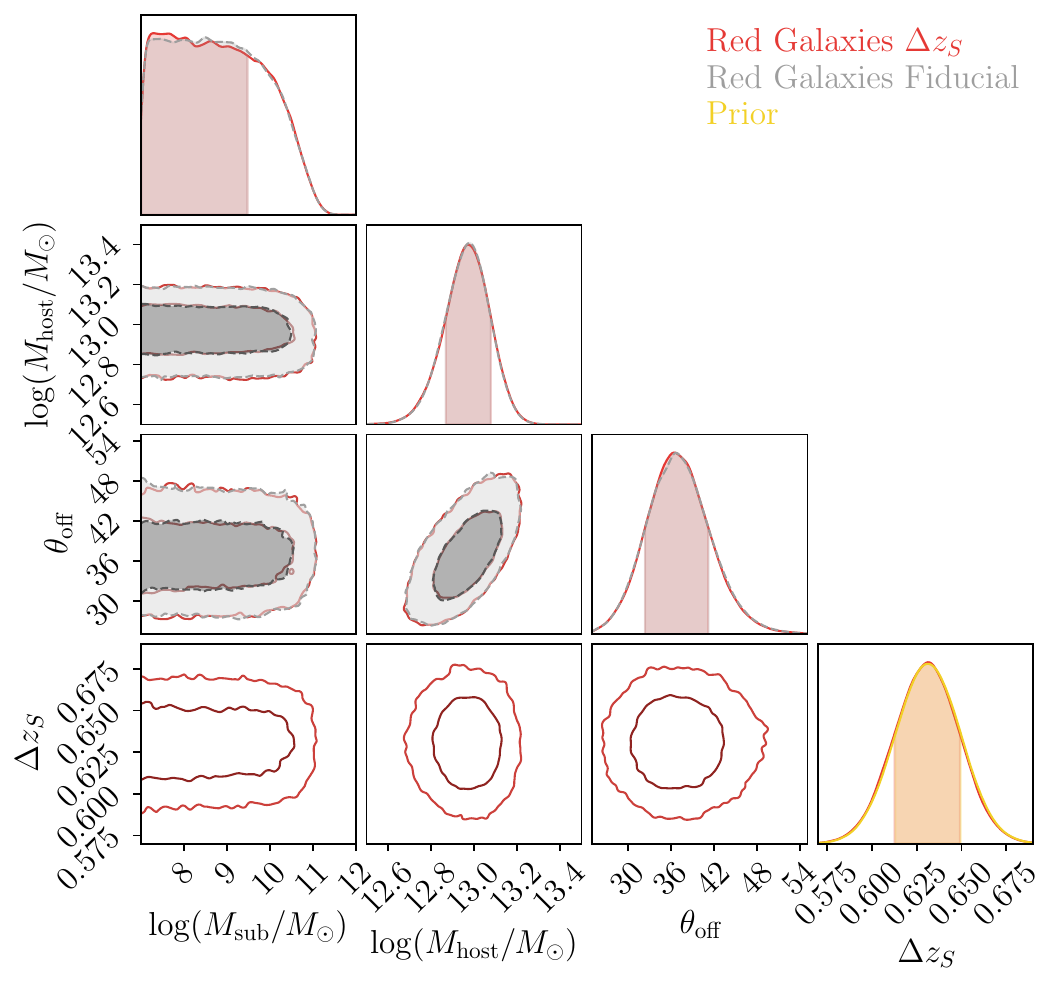}
    \caption{MCMC posterior distribution for red LSBGs coinciding with the first four rows of Table~\ref{tab:model_params_alt}. This posterior distribution includes the mean source redshift as a free parameter, constrained by $\Delta z_S$. The prior on the mean source redshift is overlaid in yellow. The fiducial model is overlaid in gray.}
    \label{fig:mcmc_sourceredshift}
\end{figure}

For (1), we include an additional free parameter $\Delta z_S$ as the shift in the source redshift and marginalize over it. 
The mean source redshift uncertainty for the individual source redshift bins are listed in \citep{Abbott2022}. We take the largest value (corresponding to bin 1) with a 2$\sigma$ of 0.018 and marginalize over it in our model. 
We find that the posterior does not change significantly, as shown in Fig.~\ref{fig:mcmc_sourceredshift}. 

\begin{figure}
    \centering
    \includegraphics[width=0.48\textwidth]{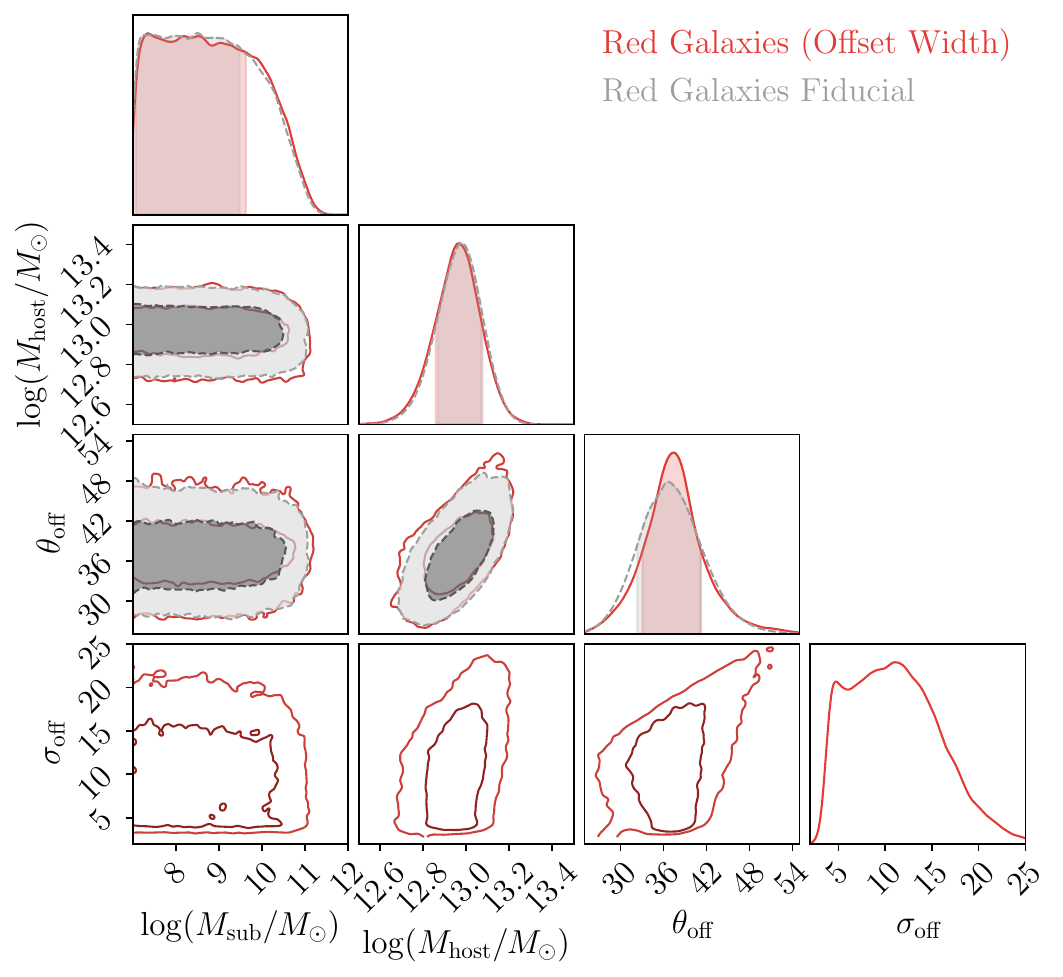}
    \caption{MCMC posterior distribution corresponding to the bottom three rows of Table~\ref{tab:model_params_alt}. The width of the offset distribution is included as a free parameter. We find that this result is consistent with the fiducial model, overlaid in gray.}
    \label{fig:mcmc_offsetwidth}
\end{figure}

For (2), we include the width of the offset distribution, $\sigma_\mathrm{off}$, as a free parameter. Note that for the fiducial model, we assume a $\sigma_\mathrm{off}$ of $12.2~\mathrm{arcmin}$, or $\frac{\theta_\mathrm{off}}{3}$. This is consistent with the posterior distribution, as shown in Fig.~\ref{fig:mcmc_offsetwidth}.

\begin{table} 
\begin{tabular}{ccc}
\hline
Parameter & Priors & Posteriors\\
\hline
 & $\Delta z_s$ Uncertainty & \\

	$M_{\rm sub}$ & U$(7, 12) \log(M_{\odot})$ & $ \log(M_\mathrm{sub}/M_\odot)< 11.50$ \\

    $M_{\rm host}$ & U$(10, 15) \log(M_{\odot})$ & $\log(M_\mathrm{host}/M_\odot) = 12.97^{+0.11}_{-0.10} $ \\ 

	$\theta_{\rm off}$ &  U$(25, 55) ~\mathrm{arcmin}$ & $36.4^{+4.8}_{-4.0}~\mathrm{arcmin}$ \\

	$\Delta z_S$ & $N(0.6312, 0.018)$ & $0.632^{+0.018}_{-0.019}$ \\	
\hline

& $\sigma_\mathrm{off}$ Included & \\

	$M_{\rm sub}$ & U$(7,  12) \log(M_{\odot})$ & $\log(M_\mathrm{sub}/M_\odot)< 11.60$ \\

    $M_{\rm host}$ & U$(10, 15) \log(M_{\odot})$ & $\log(M_\mathrm{host}/M_\odot)=12.96\pm 0.11$ \\ 

	 $\theta_{\rm off}$ &  U$(25, 55)~\mathrm{arcmin}$ & $37.4^{+3.8}_{-4.3}~\mathrm{arcmin}$ \\	

	$\sigma_{\rm off}$ & U$(2,25)~\mathrm{arcmin}$ & $\sim 12~\mathrm{arcmin}$ \\
	
\hline
\end{tabular}
\caption{Priors and mean posteriors for the red LSBG tangential shear measurements. The first three rows include the subhalo mass, host halo mass, radial offset, and source redshift as free parameters. We find that marginalizing over $\Delta z_S$ does not add to the uncertainty. These results correspond to Fig.~\ref{fig:mcmc_sourceredshift}. The bottom three rows include the width of the offset distribution as a parameter and correspond to Fig.~\ref{fig:mcmc_offsetwidth}.}
\label{tab:model_params_alt}
\end{table}

\section{Model Comparison} \label{sec:hodmodels}
In this appendix, we compare the results of our fiducial model to a model constructed under the Halo Occupation Distribution (HOD) framework, as described in \citet{Zacharegkas2022}. It is important to note that we did not select the HOD model as our fiducial because it lacks a crucial term accounting for subhalo mass, which is fundamental to this work. However, we utilize it for comparison with our host halo mass results and to assess certain assumptions of our model, such as the prevalence of red LSBG galaxies as satellites and the negligible impact of the two-halo term across the selected angular scales. In addition, rather than following a Gaussian distribution of radial offsets, the satellite galaxies in the HOD model are spatially distributed within the host halo along an NFW profile.

\begin{figure}
    \centering
    \includegraphics[width=0.45\textwidth]{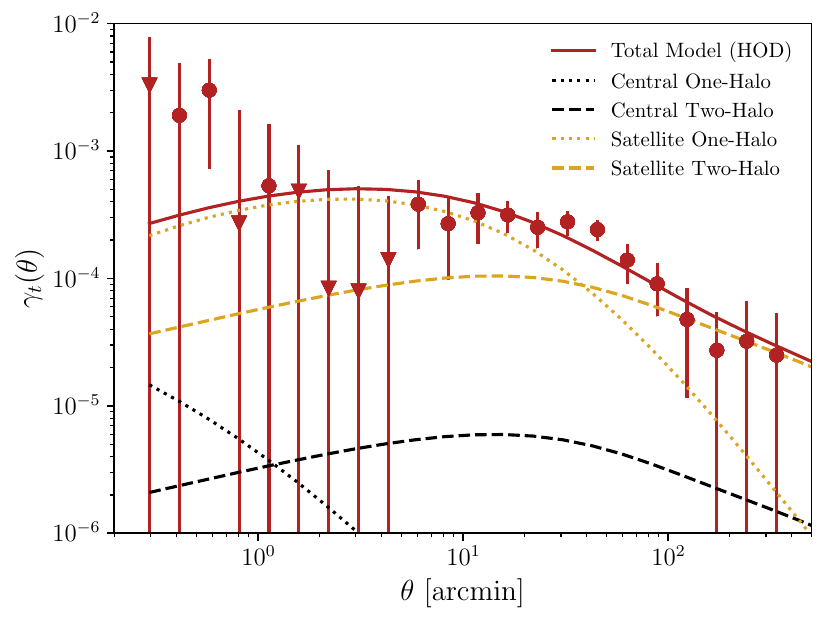}
    \caption{HOD model fit to the red LSBG tangential shear measurements, adopting the model from \citet{Zacharegkas2022}. The central one-halo and two-halo terms are subdominant and do not contribute significantly to the model. We find that the best-fit model requires a high satellite fraction of $0.94\pm 0.2$ and a host halo mass of  $\log(M_\mathrm{host}/M_\odot) = 13.1\pm0.2$. Note that the triangle markers indicate a negative shear measurement.}
    \label{fig:hod_model}
\end{figure}

 We compare the HOD model to the red LSBG shear measurements in Fig.~\ref{fig:hod_model}. We find that the HOD model for the red LSBG sample requires a satellite fraction of $0.94\pm 0.2$, validating our initial assumption of a satellite-dominated sample described in Sec.~\ref{sec:measurements}. In addition, we find that the host halo mass estimate produced by the HOD model, $\log(M_\mathrm{host}/M_\odot) = 13.1 \pm 0.2$, is consistent within 2$\sigma$ with the estimate produced by the fiducial model. We note that at larger scales, the two-halo term becomes more prominent, but we have tested that removing these scales does not significantly impact our best-fit estimate, shifting the host halo mass best fit from $\log(M_\mathrm{host}/M_\odot) = 12.98$ to $\log(M_\mathrm{host}/M_\odot) = 12.99$.
 Despite important modeling differences, the host halo mass estimate remains stable, validating our model.

\section{Lens Sample UDG Composition} \label{sec:UDGs} 

We can utilize the redshift distributions of Sec.~\ref{sec:lens catalog} and the LSBG effective radii measurements to calculate the physical size distributions of the LSBG samples and to estimate the percentage of UDGs. 
UDGs are typically defined as galaxies with a physical radius of ($R_\mathrm{eff}(g)>1.5 \mathrm{kpc}$) and a central surface brightness of $\mu_0(g)>24.0~\mathrm{mag}~\arcsec^{-2}$ (see \citet{Tanoglidis2021}). 35\% of the red and 17\% of the blue LSBGs pass the central surface brightness cut. With the caveat that these subsamples might not follow the same redshift distributions as Fig.~\ref{fig:redshift_distribution_avg}, we can estimate the physical size of these galaxies. 

\begin{figure}
    \centering
    \includegraphics[width=0.45\textwidth]{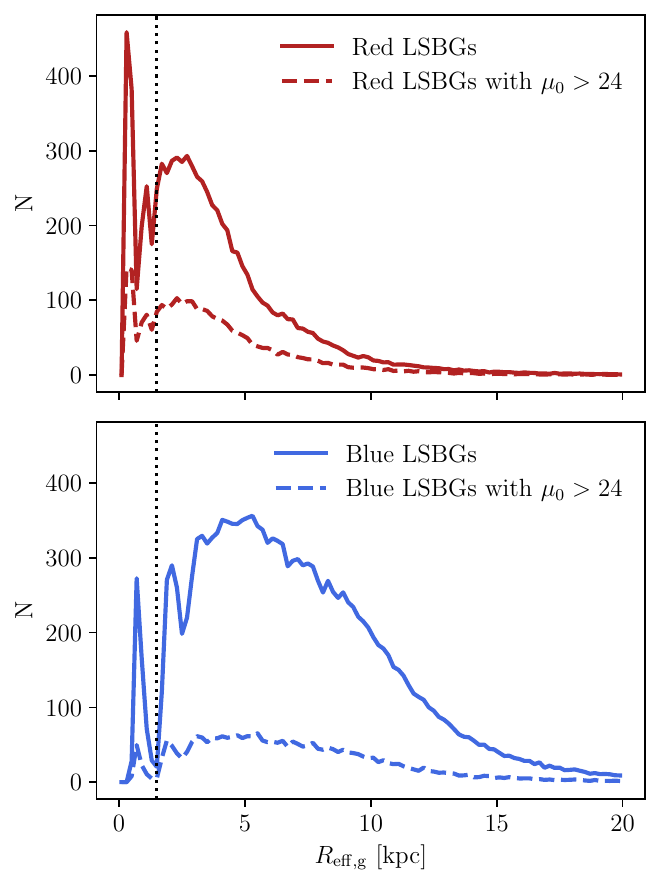}
    \caption{Physical size distribution for the red and blue LSBG samples. The dashed lines show the galaxies that meet the central surface brightness qualifications for an UDG. The black vertical dashed lines indicate the physical size cutoff for red and blue UDGs.}
    \label{fig:physical_size_distribution}
\end{figure}

 First, we convert the apparent sizes of the galaxies to an array of physical sizes corresponding to each redshift bin of the distribution. We shift from the observed angular space of the effective radii measurements to physical space by using $R_\mathrm{eff, phys} = R_\mathrm{eff, \theta}\times D_L$. We weight these physical size distributions by the normalized redshift distribution and sum the results, producing a distribution of physical sizes for both the red and blue LSBG samples: 
\begin{equation} \label{eq:physical_radius} n(R_\mathrm{eff, phys}) = \frac{\Sigma(n(R_\mathrm{eff, phys}(z_L, R_\mathrm{eff, \theta})) n(z_L)}{\Sigma n(z_L)}.
\end{equation}
Moreover, we caution that the tail in our redshift distributions can produce unphysically large galaxy sizes. 
We find that 28\% of the red and  16\% of the blue LSBG galaxies meet these conditions and can be classified as UDGs. These distributions are shown in Fig.~\ref{fig:physical_size_distribution}. 

It is interesting to note that \citet{Greene2022} found that 65-80\% of the red and 60-77\% of the blue HSC LSBG sample qualify as UDGs. However, they drew these proportions from a UDG definition solely based on size cuts, where $R_\mathrm{eff}(g)>1.5 \mathrm{kpc}$. Using this definition, we find that 71\% of our red and 89\% of our blue LSBG sample qualify as UDGs. 

\end{document}